\documentclass[11pt,a4paper]{article}
\usepackage[T1]{fontenc}
\usepackage[utf8]{inputenc}
\usepackage{newtxtext,newtxmath}
\usepackage{amsmath,bm}
\usepackage{graphicx}
\usepackage{xcolor}
\usepackage{booktabs,longtable,array,calc,tabularx}
\usepackage{caption}
\usepackage{float}
\usepackage{placeins}
\usepackage{needspace}
\usepackage{microtype}
\usepackage{geometry}
\usepackage{newunicodechar}
\usepackage{xurl}
\usepackage[hidelinks]{hyperref}

\makeatletter
\renewcommand\section{\@startsection{section}{1}{\z@}{12pt plus 2pt minus 2pt}{6pt plus 1pt minus 1pt}{\normalfont\bfseries\fontsize{14}{16}\selectfont}}
\renewcommand\subsection{\@startsection{subsection}{2}{\z@}{9pt plus 2pt minus 1pt}{2pt plus 1pt minus 1pt}{\normalfont\bfseries\fontsize{12}{14}\selectfont}}
\makeatother
\newcommand{\manuscriptheading}[1]{\par\Needspace{3\baselineskip}\noindent\textbf{#1}\par\nobreak\vspace{1pt}}

\newcommand{\tablecaption}[1]{\refstepcounter{table}\par\noindent{\small Table \thetable. #1}\par\vspace{3pt}}
\makeatletter
\def\maxwidth{\ifdim\Gin@nat@width>\linewidth\linewidth\else\Gin@nat@width\fi}
\def\maxheight{\ifdim\Gin@nat@height>0.83\textheight 0.83\textheight\else\Gin@nat@height\fi}
\makeatother
\setkeys{Gin}{keepaspectratio}

\newunicodechar{±}{\ensuremath{\pm}}
\newunicodechar{²}{\ensuremath{^{2}}}
\newunicodechar{³}{\ensuremath{^{3}}}
\newunicodechar{¹}{\ensuremath{^{1}}}
\newunicodechar{×}{\ensuremath{\times}}
\newunicodechar{≤}{\ensuremath{\leq}}
\newunicodechar{·}{\ensuremath{\cdot}}
\newunicodechar{Ł}{\L{}}
\newunicodechar{⁴}{\ensuremath{^{4}}}
\newunicodechar{⁵}{\ensuremath{^{5}}}
\newunicodechar{⁶}{\ensuremath{^{6}}}
\newunicodechar{⁷}{\ensuremath{^{7}}}
\newunicodechar{⁻}{\ensuremath{^{-}}}
\newunicodechar{−}{\ensuremath{-}}
\newunicodechar{₀}{\ensuremath{_{0}}}
\newunicodechar{₁}{\ensuremath{_{1}}}
\newunicodechar{₂}{\ensuremath{_{2}}}
\newunicodechar{₃}{\ensuremath{_{3}}}
\newunicodechar{ₓ}{\ensuremath{_{x}}}
\newunicodechar{ₛ}{\ensuremath{_{s}}}
\newunicodechar{ᵢ}{\ensuremath{_{i}}}
\newunicodechar{Δ}{\ensuremath{\Delta}}
\newunicodechar{α}{\ensuremath{\alpha}}
\newunicodechar{β}{\ensuremath{\beta}}
\newunicodechar{ε}{\ensuremath{\varepsilon}}
\newunicodechar{ζ}{\ensuremath{\zeta}}
\newunicodechar{θ}{\ensuremath{\theta}}
\newunicodechar{π}{\ensuremath{\pi}}
\newunicodechar{σ}{\ensuremath{\sigma}}
\newunicodechar{ω}{\ensuremath{\omega}}
\newunicodechar{⊙}{\ensuremath{\odot}}
\newunicodechar{⊘}{\ensuremath{\oslash}}
\newunicodechar{⟨}{\ensuremath{\langle}}
\newunicodechar{⟩}{\ensuremath{\rangle}}

\hypersetup{
  pdftitle={A hierarchy of physics-embedded neural operators for structural dynamic surrogate modeling},
  pdfauthor={Amir Nasrollahi and Mehmet Sefa Orak},
  pdfsubject={Physics-embedded neural operators for structural dynamic surrogate modeling},
  pdfkeywords={Modal Rational, PGMON, MOSAIC, structural dynamics, surrogate modeling}
}

\begin{document}

\begin{center}
{\bfseries\fontsize{17}{19}\selectfont A hierarchy of physics-embedded neural operators for structural dynamic surrogate modeling\par}
\vspace{5pt}
{\fontsize{12}{14}\selectfont Amir Nasrollahi\textsuperscript{1} and Mehmet Sefa Orak\textsuperscript{2,*}\par}
\vspace{4pt}
{\fontsize{9.5}{11}\selectfont
\textsuperscript{1} Department of Aeronautics and Astronautics, Stanford University, Stanford, CA, USA\par
\textsuperscript{2} Department of Civil Engineering, Maltepe University, Istanbul, Türkiye\par
\textsuperscript{*} Corresponding author: Mehmet Sefa Orak (\href{mailto:sefaorak@maltepe.edu.tr}{sefaorak@maltepe.edu.tr})\par}
\end{center}
\vspace{4pt}
\manuscriptheading{Abstract}

Structural-response surrogates must represent moving resonances, phase changes, and transfer channels with very different amplitudes during the optimization iteration or health state changes. In recent studies, physics-embedded machine learning techniques have been proposed to replace expensive finite-element analysis (FEA) models and achieve a comparable level of accuracy by integrating physics equations in different stages of machine learning architecture, from input data, forward propagation, loss function, and backpropagation (error optimization). In this research, we propose Modal Rational (MR), a neural mapping of structural features to modal quantities of structures followed by analytical frequency-response reconstruction. The Physics-Guided Modal Operator Network (PGMON) extends this representation through bounded frequency, damping, and residual-flexibility corrections; \textbf{M}odal \textbf{O}perator with \textbf{S}tructural GELU and \textbf{A}nchored-S\textbf{I}REN \textbf{C}orrection MOSAIC-MR and MOSAIC-PGMON, where MOSAIC denotes the bounded spectral-discrepancy correction framework used here, add bounded spectral discrepancies. These components can support finite-dimensional input--output mappings and provide a physical decoder for future neural operators. The present evidence concerns fixed-topology mappings, not a completed neural-operator comparison. Two populations, each containing 20,000 designs, are considered: an 80 m steel truss tower with variable member areas and attached masses, and a tapered shell beam with ten variable station masses. Archived response evaluations cover ten test designs and seven prescribed acceleration histories per structure. The histories are benchmark inputs only; neither example is claimed to be an earthquake-engineering design assessment. With 5\% reference modal damping, the recorded MR, PGMON, and MOSAIC configurations give mean displacement-history errors of 0.73--0.76\% for the tower and 0.029--0.116\% for the beam. These percentages are relative to the declared, band-limited modal references. Supplied modal-frequency information in the archived evaluation is distinguished from design-only deployment, and unequal parameter counts and training settings are reported explicitly. A response-error analysis separates learned modal discrepancies, truncation, and excitation weighting, explaining why acceleration and difficult tower cases show limitation and trustworthy concern of blind application of machine learning surrogates. The results support physically structured prediction while limiting claims about generalization, computational savings, and the benefit of added corrections. Developing and validating neural operators around these architectures is the next step.


\manuscriptheading{Highlights}

Modal Rational learns structural modal quantities while retaining analytical response reconstruction.\\
PGMON and MOSAIC introduce bounded, explicitly identified corrections.\\
Prescribed acceleration histories benchmark prediction; they do not establish seismic safety.\\
Response-specific analysis separates approximation error from modeling and information assumptions.

\section{1. Introduction}\label{introduction}

Repeated structural analysis appears in sizing studies, uncertainty propagation, and design exploration. The difficulty is not that one dynamic calculation is necessarily expensive; it is that many related designs and many inputs must be considered. A useful surrogate should therefore retain the distinctions that matter to a structural analyst: the design-dependent dynamics, the excitation, the measured output and the approximation used as the numerical reference. A single percentage is not enough when displacement, acceleration and peak response can react differently to the same spectral error.

For a linear structure with classical damping, the frequency response function (FRF) separates these ingredients. Its resonant denominators are determined by modal frequencies and damping, while its numerators connect modal participation to the observed response {[}1{]}. Once the FRF is available over a declared band, different excitation histories can be applied without fitting a new response model for each history. This does not mean that every time-domain learning method requires retraining for every input. Rather, the FRF construction explicitly preserves linear superposition and makes the structural and excitation dependence transparent.

Modal quantities have long been used in surrogate modeling. Pichler et al. learned modal information before reconstructing uncertain FRFs, and Gibanica and Abrahamsson developed modal-residue surrogates with treatment of omitted dynamics and modal correspondence {[}2, 3{]}. Projection-based reduced models offer a related route to inexpensive parametric dynamics {[}4{]}. The present contribution is not the modal-superposition identity itself. It is a neural formulation, Modal Rational (MR), and a controlled sequence of extensions that preserve that identity while learning the design-dependent quantities.

We distinguish physics embedded in an architecture from physics imposed through an additional differential-equation loss. Physics-informed neural networks and broader scientific machine learning provide important foundations {[}5, 6{]}. Here the principal structural prior is the analytical modal decoder. MR uses it to reconstruct the response; PGMON permits bounded changes to selected physical parameters and residual flexibility; \textbf{M}odal \textbf{O}perator with \textbf{S}tructural GELU and \textbf{A}nchored-S\textbf{I}REN \textbf{C}orrection (MOSAIC) adds a limited spectral discrepancy to either model. Additional flexibility is evaluated rather than assumed to be beneficial.

The proposed components are relevant to two learning interfaces. A vector-valued predictor can map structural inputs to nodal modal values, frequencies, or compact modal quantities. A neural operator can instead combine an input representation with output query coordinates. Deep Operator Network (DeepONet) and Fourier neural operators provide established examples of the latter approach {[}7, 8, 9{]}. The present paper develops the structural formulation and examines the archived fixed-topology evidence. Dedicated neural-operator implementations and their matched comparisons remain future work; no interim training snapshot is used as a completed experiment.

The tower and beam deliberately expose different sources of difficulty. The tower has fewer finite-element degrees of freedom but a much larger design-input space, with both stiffness and mass varying. The beam has a larger shell discretization but only ten changing mass variables. Their results are not controlled replications of one another: excitation directions, modal ordering and training settings also differ. We use the contrast to identify limitations, not to attribute every difference to mesh size or frequency bandwidth.

From a structural engineering standpoint, the key value of this work lies in bridging scientific machine learning with classical dynamics rather than deploying unconstrained black-box approximations. By directly embedding analytical modal superposition, resonant denominators, and static corrections into the neural architecture, the framework guarantees mechanical interpretability and bounded physical behavior. In practical structural engineering, this provides an efficient and trustworthy surrogate for computationally demanding multi-query tasks such as high-throughput parametric sizing, rapid seismic sensitivity screenings, and uncertainty quantification at a fraction of full finite-element cost.

\FloatBarrier
\section{2. Related work and contribution}\label{related-work-and-contribution}

\subsection{2.1. Learning structural response and modal content}\label{learning-structural-response-and-modal-content}

Rational approximation and modal surrogates both represent dynamic response through poles and associated numerators. Vector fitting provides a well-established computational foundation for rational frequency-response approximation {[}10{]}. Modal learning adds design dependence to quantities whose frequency response is known analytically. The important questions are then how those quantities are labeled, which quantities are available at inference, and how unresolved dynamics enter the prediction. The proposed MR--PGMON--MOSAIC sequence is organized around these questions rather than around activation functions alone.

Direct neural prediction of vibration and frequency responses has also been demonstrated. Schultz et al. considered a multimass oscillator, while van Delden et al. learned structural vibration responses in a broader frequency-query setting {[}11, 12{]}. Physics-guided time-history models, including the physics-guided convolutional neural network (PhyCNN), incorporate structural information through different computational paths {[}13{]}. These approaches are relevant comparators, but their input information and output tasks must be matched before attributing an accuracy difference to a particular architecture.

Learning modal fields introduces a separate correspondence problem. Eigenvector signs, mode exchanges and rotations within repeated eigenspaces can change labels without changing the physical response. Recent structural eigenvalue learning explicitly addresses modal matching {[}14{]}. Residue learning removes the single-mode sign ambiguity, but it does not remove every ambiguity caused by clustered or crossing modes. The beam preprocessing and the limits of that invariance are therefore stated explicitly below.

\subsection{2.2. Physical constraints, coordinate representations and operators}\label{physical-constraints-coordinate-representations-and-operators}

DeepXDE and physics-informed solid-mechanics models illustrate the use of data and governing equations in one learning workflow {[}15, 16{]}. Gradient-flow studies show why multiple objectives can interact poorly {[}17{]}; that literature motivates careful normalization and reporting, not an assumption that a physical loss always helps. Our modal decoder is retained independently of the choice of auxiliary supervision.

The Sinusoidal Representation Network (SIREN) and Fourier-feature networks improve the representation of rapidly varying coordinate functions {[}18, 19{]}. Spectral-bias results concern frequency in the function being approximated, not a universal preference for low physical vibration modes {[}20{]}. A narrow resonance at a low physical frequency can still be difficult to learn as a function of the query coordinate. This distinction matters when interpreting the tower errors.

SIREN and anchored SIREN remain the source study's frequency-query comparators. The anchored comparator receives modal features but does not enforce the analytical modal sum. Consequently, it helps separate the usefulness of supplied spectral information from the usefulness of an analytical decoder, subject to the unequal training and capacity budgets reported below.

Operator learning offers an extension rather than a replacement for the structural representation. Comparisons of DeepONet and Fourier neural operators emphasize the influence of features, bases and evaluation protocol {[}21{]}; error analysis distinguishes encoding, approximation and reconstruction contributions {[}22{]}. Laplace neural operators use pole--residue structure directly, while deep reduced-order models and finite-element-informed operators combine learned and reduced physical representations {[}23, 24, 25{]}. We build on these precedents without claiming that poles, modal superposition or branch--trunk learning are new concepts.

The contribution of this paper is the proposed MR formulation, its bounded PGMON and MOSAIC extensions, and a response-specific account of their archived performance. The account makes the inference information explicit, avoids treating corrections as physical guarantees, and explains why a more elaborate model need not improve every response quantity. It also identifies which conclusions require a later, fully matched operator-learning study.

\FloatBarrier
\section{3. Scope, interfaces and evidence}\label{scope-interfaces-and-evidence}

\subsection{3.1. Vector mapping and frequency-queryable prediction}\label{vector-mapping-and-frequency-queryable-prediction}

The author-defined rectified linear unit (ReLU) and Gaussian error linear unit (GELU) formulations are ordinary vector mappings. Their input contains the specified information for all nodes; their output contains modal values at every node for each predicted mode and the corresponding natural frequencies. With n nodal degrees of freedom and \emph{N} modes, the output can be written as follows. The activation choice does not itself supply a modal-response decoder {[}26, 27{]}.

\begin{equation}
\mathcal{N}_{\sigma}:\mspace{6mu}\mathbf{x}_{\text{all nodes}} \mapsto \text{[vec}\left( \widehat{\mathbf{\Phi}} \right),{\widehat{f}}_{1},\ldots,{\widehat{f}}_{N}\text{]},\quad\sigma \in \{\text{ReLU},\text{GELU}\}.
\end{equation}

\emph{where} \(\mathcal{N}_{\text{σ}}\) is the vector-valued neural mapping with activation σ; \(\mathbf{x}_{\text{all nodes}}\) is the packed all-node structural input; \(\mathbf{\Phi}\) is the predicted nodal modal-value tensor; \(f_{\text{r}}\) is the r-th predicted natural frequency; \(N\) is the number of predicted modes; \(\text{vec}\) denotes vectorization; \(\sigma\) is either the ReLU or GELU activation.

Modal Rational takes a different route as shown in Figure 1: it learns the modal quantities needed by the declared response channels, then reconstructs their frequency dependence analytically. Its neural parameter map is finite-dimensional even though the decoder can be queried continuously in frequency. PGMON and MOSAIC extend this construction. This distinction concerns the complete forward computation, not whether an internal encoder contains ReLU or GELU layers.

\begin{figure}[!htbp]
\centering
\includegraphics[width=6.2361in]{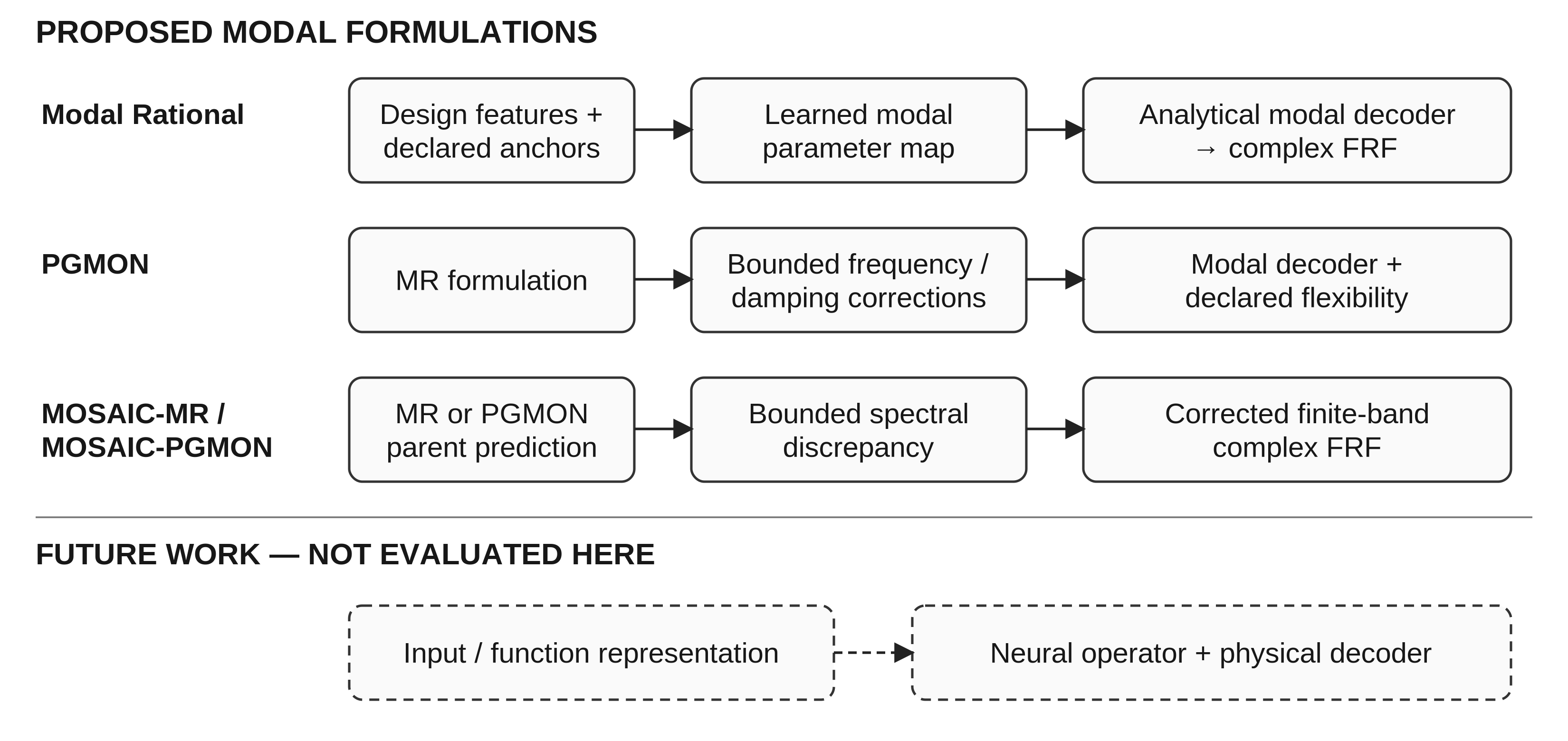}
\caption{Intended MR--PGMON--MOSAIC hierarchy and its prospective use within a neural operator. The lower, dashed extension is future work, not an additional experiment in this paper. The diagram does not prescribe a new encoder or replace an existing implementation.}
\end{figure}

\subsection{3.2. What the archived numerical results establish}\label{what-the-archived-numerical-results-establish}

The numerical results reported in this work correspond to the completed tower and beam experiments described in Sections 7 and 8. For the modal-aware configurations, the natural-frequency information specified by the corresponding dataset is supplied to the response model, while the remaining design-dependent modal quantities and admissible corrections are learned. The reported results therefore quantify the accuracy of \textbf{conditional modal-response surrogates} under a declared modal-information budget. They should not be interpreted as evidence for a fully design-only deployment in which all modal anchors are inferred from the structural design variables.

Within this information setting, Modal Rational reconstructs the complex frequency response analytically from learned, frequency-independent modal quantities. PGMON augments the same modal representation through bounded corrections to the modal frequencies, damping ratios and residual-flexibility contribution, whereas MOSAIC-MR and MOSAIC-PGMON introduce a bounded spectral discrepancy around their respective physical backbones. The numerical comparisons consequently assess complete surrogate configurations, including their input information, parameterization, auxiliary supervision and response decoder, rather than isolating the effect of an individual neural component.

The archived ReLU and GELU configurations evaluated in the completed experiments are direct spectral predictors using conventional ReLU and GELU activation functions: they map the structural design variables and query frequency to the real and imaginary components of the frequency-response outputs. By contrast, the vector-valued ReLU and GELU interface introduced in Section 3.1 predicts nodal modal quantities and natural frequencies before structural-response reconstruction. These are distinct learning tasks and are not treated as interchangeable implementations. Accordingly, the archived direct-spectral results are retained as frequency-response regression benchmarks, whereas the nodal-modal interface defines a separate model class whose performance must be established under its own training and evaluation protocol.

The reported error tables therefore support comparisons among the evaluated configurations within their stated inference conditions. They do not establish design-only accuracy, topology transfer, or superiority attributable solely to an activation function, analytical decoder, or correction branch. These broader conclusions require matched experiments in which the input information, output task, data partitions and computational budgets are held fixed.

\FloatBarrier
\section{4. Structural frequency-response formulation}\label{structural-frequency-response-formulation}

\subsection{4.1. Equations of motion and input convention}\label{equations-of-motion-and-input-convention}

Let \(\mathbf{x}\) denote one structural design. The free-coordinate displacement \(\mathbf{u}\) is measured relative to a uniformly translating support. For \(n_{\text{g}}\) excitation directions, define \(\mathbf{b}_{\text{d}}\) = (M\(\mathbf{\iota}_{\text{d}}\))f, where \(\mathbf{M}\) is the assembled mass matrix, \(\mathbf{\iota}_{\text{d}}\) represents unit rigid translation in direction \(d\), and the subscript f retains the free rows. The free-free stiffness, mass, and damping matrices are \(\mathbf{K}_{\text{ff}}\), \(\mathbf{M}_{\text{ff}}\) and \(\mathbf{C}_{\text{ff}}\). Under the uniform-support-motion assumption,

\begin{equation}
\mathbf{M}_{ff}\ddot{\mathbf{u}} + \mathbf{C}_{ff}\dot{\mathbf{u}} + \mathbf{K}_{ff}\mathbf{u} = - \sum_{d = 1}^{n_{g}}\mathbf{b}_{d}a_{d}(t),\quad\quad\mathbf{b}_{d} = \left( \mathbf{M}\mathbf{\iota}_{d} \right)_{f}.
\end{equation}

\emph{where} \(\mathbf{M}_{\text{ff}}\) is the free-free mass matrix; \(\mathbf{C}_{\text{ff}}\) is the free-free damping matrix; \(\mathbf{K}_{\text{ff}}\) is the free-free stiffness matrix; \(\mathbf{u}\) is the vector of free relative displacements; \(\mathbf{b}_{\text{d}}\) is the d-th base-influence load vector; \(a_{\text{d}}\) is the base acceleration in direction d; \(n_{\text{g}}\) is the number of excitation directions; \(\mathbf{\iota}_{\text{d}}\) is the unit rigid-translation vector in direction \(d\).

The complete mass--influence product is retained before selecting its free rows. Replacing it by \(\mathbf{M}_{\text{ff}}\) times a free-only influence vector is not generally equivalent when consistent mass couples free and restrained degrees of freedom. The scalar input \(a_{\text{d}}\) has acceleration units. Its equivalent inertial force is −\(\mathbf{b}_{\text{d}}\) \(a_{\text{d}}\); calling \(a_{\text{d}}\) itself an applied nodal force would change the physical meaning and units of the transfer function.

The formulation is restricted to linear, classically damped structural dynamics with zero initial conditions for the reconstructed histories. A nonlinear map from design variables to response does not imply geometric or material nonlinearity in an individual structure. Modal superposition follows the standard structural-dynamics construction {[}1{]}.

\begin{equation}
\mathbf{K}_{ff}\mathbf{\phi}_{r} = \omega_{r}^{2}\mathbf{M}_{ff}\mathbf{\phi}_{r},\quad\quad\mathbf{\phi}_{r}^{T}\mathbf{M}_{ff}\mathbf{\phi}_{s} = \delta_{rs},\quad\quad\Gamma_{r,d} = \mathbf{\phi}_{r}^{T}\mathbf{b}_{d}.
\end{equation}

\emph{where} \(\mathbf{\phi}_{\text{r}}\) is the r-th mass-normalized real mode shape; \(\omega_{\text{r}}\) is its natural angular frequency; \(\delta_{\text{rs}}\) is the Kronecker delta; \(\Gamma_{\text{r,d}}\) is the participation factor of mode r for input direction d.

\begin{equation}
{\ddot{q}}_{r} + 2\zeta_{r}\omega_{r}{\dot{q}}_{r} + \omega_{r}^{2}q_{r} = - \sum_{d}^{}\Gamma_{r,d}a_{d}(t).
\end{equation}

\emph{where} \(q_{\text{r}}\) is the r-th modal coordinate; \(\zeta_{\text{r}}\) is the corresponding modal damping ratio; \(\Gamma_{\text{r,d}}\) is the modal participation factor; \(a_{\text{d}}\) is the prescribed base acceleration; overdots denote time derivatives.

Natural angular frequency \(\omega_{\text{r}}\) is distinct from query angular frequency \(\nu\) = 2πf. Let \(\mathbf{L}\) select the response components and let \(o\) = (\(p\),\(c\),\(d\)) denote the response point, response component, and excitation direction. For mass-normalized real modes, the numerator of one modal contribution is

\begin{equation}
R_{o,r} = - \left( \mathbf{L}\mathbf{\phi}_{r} \right)_{p,c}\,\Gamma_{r,d},\quad\quad D_{r}(\nu) = \omega_{r}^{2} - \nu^{2} + 2i\zeta_{r}\omega_{r}\nu.
\end{equation}

\emph{where} \(R_{\text{o,r}}\) is the modal numerator for output/input channel o and mode r; \(\mathbf{L}\) is the response-selection operator; \(o\) denotes the tuple (p,c,d) of response point, component and excitation direction; \(D_{\text{r}}\) is the r-th modal denominator; \(\nu\) is the query angular frequency; \(i\) is the imaginary unit.

The product in Eq. (5) is unchanged when a mode shape changes sign, because its participation factor changes sign at the same time. With unnormalized modes, the corresponding modal contribution must be divided by the generalized mass. Omitting that factor would make the prediction depend on an arbitrary eigenvector scaling.

\begin{equation}
H_{o}(\nu) = \sum_{r \in \mathcal{I}_{N}}^{}\frac{R_{o,r}}{D_{r}(\nu)} + T_{N,o}(\nu).
\end{equation}

\emph{where} \(H_{\text{o}}\) is the complex displacement frequency-response function for channel o; \(I_{\text{N}}\) is the retained modal index set; \(T_{\text{N,o}}\) is the exact omitted-mode contribution.

The retained set need not be the first \(N\) frequency-sorted modes. \(T_{\text{N,o}}\) denotes the exact omitted contribution at this stage of the derivation. Its approximation, and the reference used to evaluate it, are separate choices rather than consequences of writing a modal sum.

\subsection{4.2. Static remainder, damping and reconstruction}\label{static-remainder-damping-and-reconstruction}

For a complete finite-element modal basis, the static response to a unit base acceleration is obtained without damping. An exact static residual for a selected modal set is

\begin{equation}
H_{0,o} = - (\mathbf{L}\mathbf{K}_{ff}^{- 1}\mathbf{b}_{d})_{p,c},\quad\quad S_{N,o} = H_{0,o} - \sum_{r \in \mathcal{I}_{N}}^{}\frac{R_{o,r}}{\omega_{r}^{2}}.
\end{equation}

\emph{where} \(H_{\text{0,o}}\) is the exact static response of channel o; \(S_{\text{N,o}}\) is the static residual associated with the retained modal set; \(\mathbf{K}_{\text{ff}}\) is the free-free stiffness matrix; \(R_{\text{o,r}}\) is the modal numerator; \(\omega_{\text{r}}\) is the natural angular frequency.

Replacing \(T_{\text{N,o}}\) by \(S_{\text{N,o}}\) restores the omitted contribution at \(\nu\) = 0, but not its resonances at nonzero frequency. The resulting expression is an approximation to the full dynamics, even when every retained modal quantity is exact. In particular, a constant remainder changes the high-frequency limit; it must not be used to claim global displacement-FRF decay or exact recovery of omitted modes.

The current reference prescription assigns \(\zeta_{\text{r}}\) = 0.05 to every mode of both structures. The earlier Rayleigh calibration, \(\mathbf{C}\) = αM + βK, instead gives \(\zeta_{\text{r}}\) = \(\alpha\)/(2\(\omega_{\text{r}}\)) + β\(\omega_{\text{r}}\)/2. Both are legitimate specified models. The present choice makes the damping assumption consistent across the two references; it is not evidence that every real mode of either physical structure has 5\% damping. Literature on damping cautions against treating a convenient numerical prescription as an experimentally identified law {[}28, 29, 30{]}.

For each input history, use its declared sampling interval and the same frequency band and taper for the reference and prediction. The upper tenth of the band is tapered continuously to zero. Writing \(A_{\text{d}}\) for the input Fourier transform, the output spectra are

\begin{equation}
U_{p,c}(\nu) = w(\nu)\sum_{d}^{}H_{p,c,d}(\nu)A_{d}(\nu),\quad\quad V_{p,c} = i\nu U_{p,c},\quad\quad A_{p,c}^{\text{rel}} = - \nu^{2}U_{p,c}.
\end{equation}

\emph{where} \(w\) is the real-valued frequency-band taper; \(A_{\text{d}}\) is the Fourier transform of input acceleration d; \(U_{\text{p,c}}\) is the relative-displacement spectrum; \(V_{\text{p,c}}\) is the relative-velocity spectrum; \(A_{\text{p,c}}^{\text{rel}}\) is the relative-acceleration spectrum.

\begin{equation}
A_{p,c}^{\text{abs}}(\nu) = A_{p,c}^{\text{rel}}(\nu) + w(\nu)\sum_{d}^{}I_{c,d}A_{d}(\nu).
\end{equation}

\emph{where} \(A_{\text{p,c}}^{\text{abs}}\) is the absolute-acceleration spectrum; \(I_{\text{c,d}}\) selects the base-acceleration direction associated with measured component c; \(A_{\text{d}}\) is the input-acceleration spectrum.

Here \(I_{\text{c,d}}\) selects the input translation associated with the measured component. The direct acceleration term is added once, with the same filter used in the relative response. Inverse transforms give the histories on the common retained time window. The reported errors are therefore finite-band reconstruction errors, not certification of an unfiltered response at arbitrary frequency.

Zero-padding reduces circular wrap-around, but its adequacy depends on the decay time of the system; doubling a signal length is not a universal guarantee. A real inverse fast Fourier transform (FFT) also imposes self-conjugate endpoint conventions. These numerical conventions must be distinguished from a physical statement about the continuous FRF. Existing results are retained under their archived reconstruction settings, and unreported padding or endpoint checks are not assumed to have passed.

The band edge is an evaluation choice, not the highest stored natural frequency. Modes above the working band may influence static and off-resonant response. Conversely, retaining a fixed number of modes does not establish that every important resonance inside the band has been included. These distinctions are especially important for the tower analysis in Section 9.

\FloatBarrier
\section{5. Proposed neural formulations}\label{proposed-neural-formulations}

\subsection{5.1. Modal Rational}\label{modal-rational}

Modal Rational (MR) is the first proposed formulation, not an external reference solution. A neural map predicts frequency-independent modal quantities, and an analytical decoder supplies their dependence on query frequency. The formulation accommodates a train-fitted regression component followed by a learned residual. It does not depend on a particular activation, hidden width, or software class name.

\begin{equation}
\widehat{\mathbf{\theta}}\left( \mathbf{x} \right) = \mathbf{\theta}^{\text{fit}}\left( \widetilde{\mathbf{x}} \right) + \mathbf{s}_{\theta} \odot \mathcal{N}_{\theta}\left( \widetilde{\mathbf{x}} \right),\quad\quad\widetilde{\mathbf{x}} = \left( \mathbf{x} - \mathbf{\mu}_{x} \right) \oslash \mathbf{s}_{x}.
\end{equation}

\emph{where} \(\mathbf{\theta}\) is the vector of predicted frequency-independent modal quantities; \(\mathbf{\theta}_{\text{fit}}\) is the train-fitted component; \(\mathbf{s}_{\text{θ}}\) is the stored output scale; \(\mathbf{N}_{\text{θ}}\) is the learned residual map; \(\mathbf{x}\) is the standardized design vector; \(\mathbf{\mu}_{\text{x}}\) is the training-set input mean; \(\mathbf{s}_{\text{x}}\) is the training-set input scale; \(\text{⊙}\) denotes elementwise multiplication; \(\text{⊘}\) denotes elementwise division.

The symbols \(\text{⊙}\) and \(\text{⊘}\) denote elementwise multiplication and division. The vector \(\mathbf{\theta}\) collects only the quantities emitted by the actual implementation, such as modal numerators and a static remainder. The fitted map may be a constant mean or the regression component used by the selected implementation. Its coefficients and the normalization statistics are determined on the training split. Equation (10) specifies the parameterization, not a claim that every archived implementation used the same fitted features.

\begin{equation}
{\widehat{H}}_{\text{MR},o}\left( \mathbf{x},\nu \right) = \sum_{r \in \mathcal{I}_{N}}^{}\frac{{\widehat{R}}_{o,r}\left( \mathbf{x} \right)}{\left( \omega_{r}^{a} \right)^{2} - \nu^{2} + 2i\zeta_{r}^{a}\omega_{r}^{a}\nu} + {\widehat{S}}_{o}\left( \mathbf{x} \right).
\end{equation}

\emph{where} \(H_{\text{MR,o}}\) is the Modal Rational prediction for channel o; \(R_{\text{o,r}}\) is the predicted modal numerator; \(\omega_{\text{r}}^{\text{a}}\) is the declared anchor frequency; \(\zeta_{\text{r}}^{\text{a}}\) is the declared anchor damping ratio; \(S_{\text{o}}\) is the declared static or residual contribution; \(\mathbf{x}\) is the structural design; \(\nu\) is query angular frequency.

The superscript a denotes declared anchor information. In a design-only implementation, the frequencies must be predicted from \(\mathbf{x}\) by a documented component of the deployed model. In the archived tower evaluation, the source formulation supplies the design's modal frequencies from its database. These are different information budgets. The error tables in this paper do not certify that substituting predicted anchors would leave the reported response errors unchanged.

MR avoids asking an unconstrained network to recreate each resonant denominator. The predicted numerator carries the output amplitude and participation, while the modal denominator determines the basic frequency dependence. This is the proposed architectural contribution; it builds on, rather than replaces, classical modal-surrogate and rational-approximation principles {[}2, 3, 10{]}. Compact and larger neural implementations belong to the same formulation only when their actual forward computations satisfy that definition.

The archived source also describes a zero-initialized bounded spectral correction for some entries labeled MR. Such a correction is not invisible: it changes the deployed model and is accounted for with the discrepancy construction in Section 5.3. Initializing that branch at zero recovers the predicted modal core, not necessarily the exact reference response. The association between those historical entries and the selected compact MR implementation must be preserved in the experiment manifest rather than resolved by changing a name.

\subsection{5.2. Common neural components}\label{common-neural-components}

Residual encoders use standard skip connections {[}31{]}; sinusoidal components use the Sinusoidal Representation Networks (SIREN) initialization and periodic activations {[}18{]}. ReLU and GELU are activation choices inside neural maps, not definitions of modal operators.

\subsection{5.3. PGMON and the MOSAIC extensions}\label{pgmon-and-the-mosaic-extensions}

PGMON extends MR by permitting bounded changes to its frequencies, damping, and declared residual flexibility. With raw head outputs \(\mathbf{z}\), the frequency and damping maps in the source formulation are

\begin{equation}
{\widehat{\omega}}_{r} = \omega_{r}^{a}\left( 1 + \beta_{\max}\tanh\left( z_{r}^{\omega} \right) \right),\quad\quad{\widehat{\zeta}}_{r} = \zeta_{\min} + \left( \zeta_{\max} - \zeta_{\min} \right)\text{sigmoid}\left( z_{r}^{\zeta} \right).
\end{equation}

\emph{where} \(\omega_{\text{r}}\) is the corrected modal frequency; \(\zeta_{\text{r}}\) is the corrected modal damping ratio; \(\omega_{\text{r}}^{\text{a}}\) is the frequency anchor; \(\beta_{\text{max}}\) is the maximum relative frequency correction; \(z_{\text{r}}^{\text{ω}}\) is the raw frequency-correction output; \(z_{\text{r}}^{\text{ζ}}\) is the raw damping output; \(\zeta_{\text{min}}\) is the lower damping bound; \(\zeta_{\text{max}}\) is the upper damping bound; \(\text{tanh}\) denotes the hyperbolic-tangent function; \(\text{sigmoid}\) denotes the logistic function.

The reported values are \(\beta_{\text{max}}\) = 0.03, \(\zeta_{\text{min}}\) = 0.005 and \(\zeta_{\text{max}}\) = 0.30. The 5\% reference damping is inside this range. The frequency head predicts a correction to its anchor, not an independent exact frequency. With supplied exact anchors, any learned shift must be interpreted as part of the approximation, potentially compensating for other discrepancies.

\begin{equation}
{\widehat{H}}_{\text{PGMON},o} = \sum_{r}^{}\frac{{\widehat{R}}_{o,r}}{{\widehat{\omega}}_{r}^{2} - \nu^{2} + 2i{\widehat{\zeta}}_{r}{\widehat{\omega}}_{r}\nu} + {\widehat{T}}_{o}(\nu).
\end{equation}

\emph{where} \(H_{\text{PGMON,o}}\) is the PGMON prediction for channel o; \(R_{\text{o,r}}\) is the modal numerator; \(\omega_{\text{r}}\) is the corrected modal frequency; \(\zeta_{\text{r}}\) is the corrected damping ratio; \(T_{\text{o}}\) is the declared residual-flexibility contribution.

The residual term must be defined by the actual configuration. The archived tower specification uses the finite-band polynomial below, with coefficients limited by its declared scale. This polynomial is a flexible discrepancy representation; it is not automatically an exact static correction or a globally proper transfer function.

\begin{equation}
{\widehat{T}}_{o}(\nu) = \left( c_{o,0} + ic_{o,1} \right) + (\nu/\Omega)^{2}\left( c_{o,2} + ic_{o,3} \right),\quad\quad c_{o,k} = c_{\text{flex}}\tanh\left( z_{o,k}^{c} \right).
\end{equation}

\emph{where} \(c_{\text{o,k}}\) are the finite-band flexibility coefficients; \(c_{\text{flex}}\) is their prescribed bound; \(z_{\text{o,k}}^{\text{c}}\) is the corresponding raw network output; \(\Omega\) equals 2πf\_max and normalizes the query frequency; \(f_{\text{max}}\) is the upper analysis-band frequency.

A complex constant can violate the real static limit, and the quadratic term does not have the modal displacement decay at large frequency. We therefore retain a finite-band approximation claim. No passivity or complete-model causality guarantee is inferred from a well-fitted polynomial. A real static contribution and an incremental flexibility correction must also not be counted twice.

MOSAIC-MR and MOSAIC-PGMON add a bounded learned spectral discrepancy to the named parent. In standardized response coordinates, the gated form is

\begin{equation}
\widehat{\mathbf{z}} = {\widehat{\mathbf{z}}}_{F} + \mathbf{\gamma} \odot \alpha\tanh\left( \mathbf{d} \right),\quad\quad\mathbf{\gamma} = \text{sigmoid}\left( \mathbf{g} \right),\quad F \in \{\text{MR},\text{PGMON}\}.
\end{equation}

\emph{where} \(\mathbf{z}\) is the corrected standardized response; \(\mathbf{z}_{\text{F}}\) is the standardized response of parent F; \(\mathbf{\gamma}\) is the learned gate; \(\alpha\) is the discrepancy amplitude; \(\mathbf{d}\) is the discrepancy-network output; \(\mathbf{g}\) is the raw gate output; \(F\) is either MR or PGMON.

The real and imaginary components are handled as separately standardized outputs. Multiplying a standardized correction by the stored scale adds a physical increment; the output mean must not be added again. A componentwise tanh bound controls each component, not the complex modulus by the same number. If both components have bound a, the modulus is bounded by \(\sqrt{2}a\).

The tower appendix reports an ungated correction with amplitude 0.03 in standardized units, whereas the beam configuration uses a logistic gate and a 0.05 local-scale correction. These configurations are stated separately; the numerical comparison is not evidence that gating alone causes their differences. Zero discrepancy recovers the parent prediction. Positive modal frequencies and damping stabilize the rational core, but an arbitrary frequency-domain correction does not automatically inherit all of the core's dynamical properties.

\subsection{5.4. Beam-specific parameterization and physical limits}\label{beam-specific-parameterization-and-physical-limits}

For the beam population, the design-to-parameter map is enriched with feature-wise linear modulation (FiLM), attention-based modal tokens, and parallel residue pathways. FiLM applies a design-dependent affine modulation to the spectral features {[}32{]}, whereas self-attention models interactions among modal tokens constructed from normalized modal frequencies and adjacent frequency gaps {[}33{]}. These additions increase the expressiveness of the learned parameterization without modifying the governing modal equations or the benchmark definition. The forward model assigns each nonmodal contribution a distinct role: the retained modal sum is complemented by one explicit static term and four structured residual poles placed above the analysis band; PGMON additionally introduces the prescribed finite-band flexibility correction, while MOSAIC adds a single bounded spectral discrepancy to the completed parent response. Each contribution is evaluated once, thereby avoiding double counting among the static remainder, residual-pole approximation, flexibility correction, and learned spectral discrepancy.

\begin{equation}
\text{FiLM}\left( \mathbf{h},\mathbf{z} \right) = \left( 1 + 0.1\tanh\left( \mathbf{\gamma} \right) \right) \odot \text{LN}\left( \mathbf{h} \right) + 0.1\tanh\left( \mathbf{\beta} \right).
\end{equation}

\emph{where} \(\mathbf{h}\) is the feature vector to be modulated; \(\mathbf{z}\) is the design latent vector; \(\mathbf{\gamma}\) is the learned scale vector; \(\mathbf{\beta}\) is the learned shift vector; \(\text{LN}\) denotes layer normalization; \(\text{⊙}\) denotes elementwise multiplication.

The residue branch factorizes response and excitation dependence mode by mode. A representative expression for the complete parameter map is

\begin{equation}
{\widehat{\mathbf{R}}}_{r} = \mathbf{l}_{r}\mathbf{b}_{r}^{T} + \Delta\mathbf{R}_{r}^{\text{latent}} + \mathbf{R}_{r}^{\text{fit}}.
\end{equation}

\emph{where} \(R_{\text{r}}\) is the complete residue matrix for mode r; \(\mathbf{l}_{\text{r}}\) is its output factor; \(\mathbf{b}_{\text{r}}\) is its excitation factor; \({\Delta\mathbf{R}}_{\text{latent}}\) is a dense latent residual; \(\mathbf{R}_{\text{fit}}\) is the fitted residue addition.

Only the first term in Eq. (17) is necessarily rank one. Dense residuals, output-dependent de-standardization and fitted additions can remove that property from the complete residue. Accordingly, the beam implementation is described as using a factorized branch, not as enforcing an exact rank-one residue matrix at every mode. Any stronger property would have to be checked on the final physical tensor.

The source also includes four structured residual poles above the band edge, an explicit static contribution and the declared polynomial correction. In a rational residual branch,

\begin{equation}
T_{o}^{\text{rat}}(\nu) = \sum_{k = 1}^{K}\frac{A_{o,k}}{\Omega_{k}^{2} - \nu^{2} + 2i\eta_{k}\Omega_{k}\nu},\quad\quad f_{k} = f_{\max} + \text{softplus}\left( p_{k} \right),\quad\Omega_{k} = 2\pi f_{k}.
\end{equation}

\emph{where} \(T_{\text{o}}^{\text{rat}}\) is the rational residual contribution; \(A_{\text{o,k}}\) is the k-th residual-pole numerator; \(\Omega_{\text{k}}\) is its angular frequency; \(\eta_{\text{k}}\) is its damping ratio; \(K\) is the number of residual poles; \(f_{\text{k}}\) is the residual-pole frequency; \(p_{\text{k}}\) is its raw parameter; \(f_{\text{max}}\) is the analysis-band edge; \(\text{softplus}\) enforces a positive offset above the band.

Real numerators and positive \(\eta_{\text{k}}\), \(\Omega_{\text{k}}\) give stable second-order terms. Poles constrained above the band may approximate an off-resonant remainder, but they do not explicitly restore a missing physical pole inside the band. An accurate static value is therefore not sufficient evidence that the retained dynamics are complete.

\Needspace{7\baselineskip}
\begin{table}[H]
\tablecaption{Reported size and configuration differences. Counts are millions of trainable parameters; they are not matched across methods.}

\centering
\begin{tabular}{@{}
  >{\raggedright\arraybackslash}p{(\columnwidth - 4\tabcolsep) * \real{0.6032}}
  >{\raggedright\arraybackslash}p{(\columnwidth - 4\tabcolsep) * \real{0.1984}}
  >{\raggedright\arraybackslash}p{(\columnwidth - 4\tabcolsep) * \real{0.1984}}@{}}
\toprule\noalign{}
\begin{minipage}[b]{\linewidth}\raggedright
\textbf{Method}
\end{minipage} & \begin{minipage}[b]{\linewidth}\raggedright
\textbf{Tower}
\end{minipage} & \begin{minipage}[b]{\linewidth}\raggedright
\textbf{Beam}
\end{minipage} \\
\midrule\noalign{}
ReLU* & 3.52 & 8.11 \\
GELU* & 3.52 & 8.11 \\
SIREN & 2.73 & 8.11 \\
Anchored SIREN & 3.76 & 8.15 \\
Modal Rational & 3.00 & 3.57 \\
PGMON & 3.22 & 3.58 \\
MOSAIC-MR & 3.20 & 9.33 \\
MOSAIC-PGMON & 3.41 & 9.33 \\
\bottomrule\noalign{}
\end{tabular}
\end{table}

\emph{* Implementation labels. These parameter counts and the corresponding response rows do not establish that the author-specified nodal-modal baseline in Eq. (1) was the implementation evaluated. The intended compact MR and the historical MR configuration likewise require source-level association; no count or error has been reassigned.}

The SIREN and anchored families use different encoder paths and activations, and the beam uses richer losses and conditioning than the tower. We therefore report practical configuration comparisons. We do not claim an equal-parameter ablation or that one added component alone explains the response differences.

\FloatBarrier
\section{6. Training and model selection}\label{training-and-model-selection}

\subsection{6.1. Targets and normalization}\label{targets-and-normalization}

Modal databases supply training targets and auxiliary quantities. Each design belongs to one fixed training, validation or test partition. Means, scales, and fitted regression components are estimated from training designs only. A small output scale requires particular care: normalizing a nearly zero channel by an arbitrarily small number can turn numerical noise into a dominant loss. The beam configuration reports an inactive-output mask based on training-target variability; its effect must be recorded separately from physical symmetry and from the number of output channels.

\begin{equation}
z_{q,o} = \frac{y_{q,o} - \mu_{o}}{s_{o}},\quad\quad{\widehat{y}}_{q,o} = \mu_{o} + s_{o}{\widehat{z}}_{q,o},\quad\quad\Delta{\widehat{y}}_{q,o} = s_{o}\Delta{\widehat{z}}_{q,o}.
\end{equation}

\emph{where} \(z_{\text{q,o}}\) is the standardized output at query q and channel o; \(y_{\text{q,o}}\) is the corresponding physical output; \(\mu_{\text{o}}\) is the training-set output mean; \(s_{\text{o}}\) is the training-set output scale; \({\Delta y}_{\text{q,o}}\) is the physical increment; \({\Delta z}_{\text{q,o}}\) is the standardized increment.

For the archived FRF-output configurations, each complex target is represented by its real and imaginary parts. Modal heads use their own stored training statistics. Training and evaluation must invert the same transformations. Neither the number of retained modes nor the number of frequencies is a reason to silently change an averaging convention.

\subsection{6.2. Frequency sampling and objectives}\label{frequency-sampling-and-objectives}

The tower draws 60\% of queries near a design's modal frequencies and the remainder uniformly over the band. The beam uses resonance, antiresonance, uniform, logarithmic, band-edge, and response-energy components. These are not identical sampling protocols. Applying a further frequency weight changes the effective objective unless an importance correction is included; the present results are described under their actual weighted recipes, not as an unbiased uniform-band integral.

\begin{equation}
\mathcal{L =}\sum_{k}^{}\lambda_{k}\mathcal{L}_{k},\quad\quad\mathcal{L}_{\text{MSE}} = \frac{1}{B\, n_{z}}\sum_{b = 1}^{B}{\sum_{j = 1}^{n_{z}}W_{b}}\left( {\widehat{z}}_{b,j} - z_{b,j} \right)^{2}.
\end{equation}

\emph{where} \(\mathcal{L}\) is the total training objective; \(\mathcal{L}_{\text{k}}\) is its k-th component; \(\lambda_{\text{k}}\) is the corresponding loss weight; \(B\) is batch size; \(n_{\text{z}}\) is the number of standardized scalar output components; \(W_{\text{b}}\) is the sample weight; \({\widehat{z}}_{\text{b,j}}\) and z\_b,j are predicted and reference standardized components.

Here \(n_{\text{z}}\) counts the standardized scalar output components. The tower objective combines standardized mean-squared error (MSE), complex relative error, log-magnitude and phase terms. The beam additionally uses derivative, peak and response-consistency terms, together with the active modal supervision and correction penalties. The squared complex magnitude is evaluated as the sum of squared real and imaginary components. Phase terms require explicit near-zero safeguards even when they avoid an angle function.

Auxiliary supervision can supply information not available to a pure output-fit comparator. For example, a residue-supervised model receives labeled modal content in addition to response samples. This is a legitimate design choice, but it prevents interpreting every improvement as an effect of the decoder alone. The complete training objective, rather than only the FRF loss, must be associated with each archived checkpoint.

Both structures use the Adam optimizer with decoupled weight decay (AdamW), with the adaptive-moment mechanism and decoupled weight decay distinguished from one another {[}34, 35{]}. Dropout is used as reported in the training configuration {[}36{]}. The beam additionally uses exponential weight averaging, implemented through an exponential moving average (EMA), and staged activation of objectives. No new optimizer or training run was introduced for this manuscript revision. Exact settings retained from the source are summarized in Appendix A.

\subsection{6.3. Checkpoint selection and evaluation separation}\label{checkpoint-selection-and-evaluation-separation}

Model selection is performed using the validation set, whereas the final response metrics are evaluated on the test designs. In addition to selecting the epoch with the lowest validation metric, the source study considered a gap-penalized score that accounts for the difference between deterministic training and validation measurements:

\begin{equation}
J(e) = V(e) + \lambda max\{ 0,V(e) - T(e)\}.
\end{equation}

\emph{where} \(J(e)\) is the gap-penalized checkpoint score at epoch e; \(V(e)\) is the deterministic validation metric; \(T(e)\) is the matched deterministic training metric; \(\lambda\) is the nonnegative gap-penalty weight; \(e\) is the epoch index.

For this comparison, \(T\) and \(V\) must be computed using the same metric, inference mode, and fixed sampling convention. A stochastic minibatch loss evaluated with dropout enabled is therefore not directly comparable with a deterministic training-set probe. The gap-penalized score is used only as an alternative checkpoint-selection diagnostic and does not constitute an independent estimate of generalization performance. Moreover, repeated model-development decisions based on the same validation set may still introduce selection bias {[}37{]}.

Only SIREN yielded different checkpoint selections under the two criteria: epoch 299 minimized the validation metric, whereas epoch 295 minimized the gap-penalized score, with corresponding displacement errors of 4.54\% and 4.53\%. The anchored and modal configurations selected the same checkpoint under both criteria. This agreement does not imply that the training--validation gap is negligible; it indicates only that, for these archived tower runs, including the gap penalty did not materially alter checkpoint selection or the reported displacement response.

\Needspace{7\baselineskip}
\begin{table}[H]
\tablecaption{Archived tower checkpoint-selection observation. The displacement metric is a response-history error, not the composite training loss.}

\centering
\begin{tabular}{@{}
  >{\raggedright\arraybackslash}p{(\columnwidth - 6\tabcolsep) * \real{0.3492}}
  >{\raggedright\arraybackslash}p{(\columnwidth - 6\tabcolsep) * \real{0.1984}}
  >{\raggedright\arraybackslash}p{(\columnwidth - 6\tabcolsep) * \real{0.1984}}
  >{\raggedright\arraybackslash}p{(\columnwidth - 6\tabcolsep) * \real{0.2540}}@{}}
\toprule\noalign{}
\begin{minipage}[b]{\linewidth}\raggedright
\textbf{Method}
\end{minipage} & \begin{minipage}[b]{\linewidth}\raggedright
\textbf{Lowest-V epoch}
\end{minipage} & \begin{minipage}[b]{\linewidth}\raggedright
\textbf{Gap-score epoch}
\end{minipage} & \begin{minipage}[b]{\linewidth}\raggedright
\textbf{Displacement error (\%)}
\end{minipage} \\
\midrule\noalign{}
SIREN & 299 & 295 & 4.54 / 4.53 \\
Anchored SIREN & 270 & 270 & 0.72 / 0.72 \\
Modal Rational & 279 & 279 & 0.73 / 0.73 \\
PGMON & 271 & 271 & 0.74 / 0.74 \\
MOSAIC-MR & 279 & 279 & 0.76 / 0.76 \\
MOSAIC-PGMON & 271 & 271 & 0.73 / 0.73 \\
\bottomrule\noalign{}
\end{tabular}
\end{table}

Identical selections do not show that a large train--validation gap is harmless. If \(T(e)\) = c\(V(e)\) with a constant 0 ≤ \(c\) \textless{} 1, then \(J(e)\) = {[}1 + \(\lambda\)(1 − \(c\)){]}\(V(e)\), and the two criteria have the same minimizer by construction. The useful observation is narrower: the additional rule did not materially change the reported tower responses for these runs.

The comparisons use one training seed per structural example. Their ten test designs, seven inputs, and multiple correlated response channels do not provide independent training-seed replicates. We therefore report observed means and maxima without converting small differences into statistical superiority claims. An observed maximum is not a certified worst case over the design population.

The response tables and Figure 2 use the completed runs. Later 3,000 design validation snapshots are not substituted for them: early-epoch FRF errors, final response-history errors and composite losses measure different quantities. Reproducibility requires each result to retain its checkpoint, source version and evaluation protocol.

\subsection{6.4. Convergence evidence and checkpoint interpretation}\label{convergence-evidence-and-checkpoint-interpretation}

Figure 2 restores the archived convergence evidence. The tower panel reports composite validation loss, whereas the beam panel reports a validation response error. The axes therefore do not measure the same quantity. Even within the tower panel, different objectives prevent treating a smaller ordinate as a cross-method accuracy ranking. Those rankings must come from the common response evaluation, not from differently weighted training objectives.

\begin{figure}[!htbp]
\centering
\includegraphics[width=6.2635in]{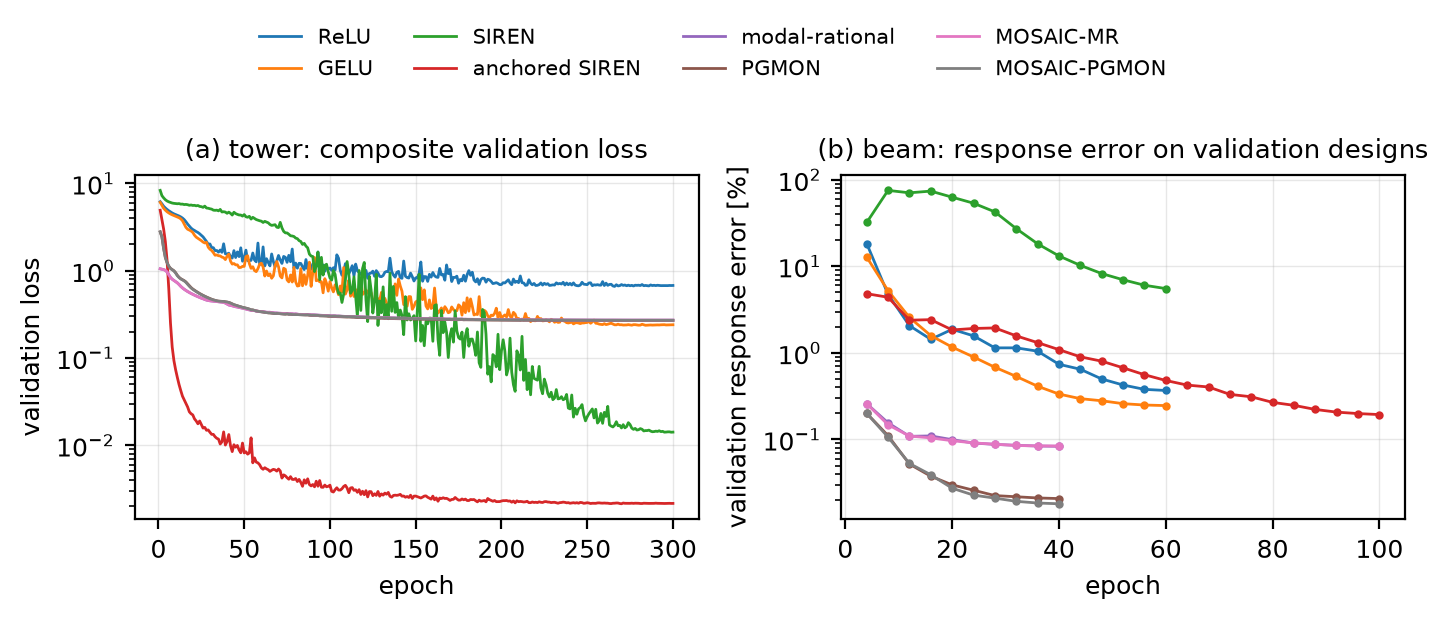}
\caption{Archived learning histories restored from the source: (a) tower composite validation loss and (b) beam validation response error. The legend is outside the data areas. Training lengths and objectives differ, so curve heights and endpoint epochs are not directly comparable across tasks. These histories belong to the completed study, not the later ongoing campaign.}
\end{figure}

The different beam endpoints reflect the reported 40--100-epoch training budgets. They are not a controlled time-to-convergence experiment. Likewise, a continuing fall in training loss need not imply an equal reduction in validation response error. The checkpoint-selection observation in Table 2 remains a separate diagnostic: it records which epoch each rule selected, rather than proving that a large training--validation gap is harmless.

\FloatBarrier
\section{7. Numerical examples and benchmark protocol}\label{numerical-examples-and-benchmark-protocol}

For this research, we developed a code for automatic

\Needspace{7\baselineskip}
\begin{table}[H]
\tablecaption{Structural populations and the archived response-evaluation tasks.}

\centering
\begin{tabular}{@{}
  >{\raggedright\arraybackslash}p{(\columnwidth - 4\tabcolsep) * \real{0.3333}}
  >{\raggedright\arraybackslash}p{(\columnwidth - 4\tabcolsep) * \real{0.3333}}
  >{\raggedright\arraybackslash}p{(\columnwidth - 4\tabcolsep) * \real{0.3333}}@{}}
\toprule\noalign{}
\begin{minipage}[b]{\linewidth}\raggedright
\textbf{Quantity}
\end{minipage} & \begin{minipage}[b]{\linewidth}\raggedright
\textbf{Steel truss tower}
\end{minipage} & \begin{minipage}[b]{\linewidth}\raggedright
\textbf{Tapered box beam}
\end{minipage} \\
\midrule\noalign{}
Elements & 582 two-node axial members & 4,000 four-node shell elements \\
Nodes / reported degrees of freedom (DOFs) & 153 / 429 free translations & 4,040 / 24,240 total nodal DOFs¹ \\
Design inputs & 32 area factors + 153 mass coordinates & 10 station masses \\
Designs; train / validation / test & 20,000; 14,000 / 3,000 / 3,000 & 20,000; 14,000 / 3,000 / 3,000 \\
Retained modal representation & 20 frequency-ordered modes & 9 participation-selected modes \\
Saved modal information & Complete 429-mode database & 20 computed modes and static data \\
Excitation / response channels & X input; 5 X-response outputs & 3 inputs × 3 centroid outputs \\
Query band & 0--7 Hz & 0--100 Hz \\
Reference modal damping & 5\% for every mode & 5\% for every mode \\
Completed history evaluation & 10 test designs × 7 signals & 10 test designs × 7 signals \\
\bottomrule\noalign{}
\end{tabular}
\end{table}

\emph{¹ The source reports 24,240 degrees of freedom for 4,040 six-DOF shell nodes. That product is a total nodal count; the number remaining after clamping is a distinct quantity and is not inferred here. The table does not label 24,240 as an independently verified free-DOF count.}

Both structural populations retain a fixed topology and discretization across all design realizations. The present study therefore evaluates generalization over parametric variations within a common finite-element model, rather than transfer across different meshes or structural topologies. Although the use of compact design variables and modal descriptors decouples the surrogate input dimension from the number of finite-element nodes, this architectural property does not establish mesh independence, discretization convergence, or transferability to a different finite-element model. For the beam, 4,040 six-degree-of-freedom shell nodes correspond to 24,240 total nodal degrees of freedom before accounting for boundary constraints; the number of unconstrained degrees of freedom is distinct and is not inferred from this product.

\subsection{7.1. Steel truss tower}\label{steel-truss-tower}

The tower is an approximately 80 m pin-jointed structure with 153 nodes, 582 axial members and 32 area groups. Ten base nodes are fixed, leaving 429 translational degrees of freedom. Its lower decagonal portion is 32 m high and approximately 20 m across; the upper hexagonal portion is 48 m high and approximately 12 m across. We use the physical height rather than the historical project label ``26-story.'' The tower geometry is shown in Figure 3.

\begin{figure}[!htbp]
\centering
\includegraphics[width=4.9771in]{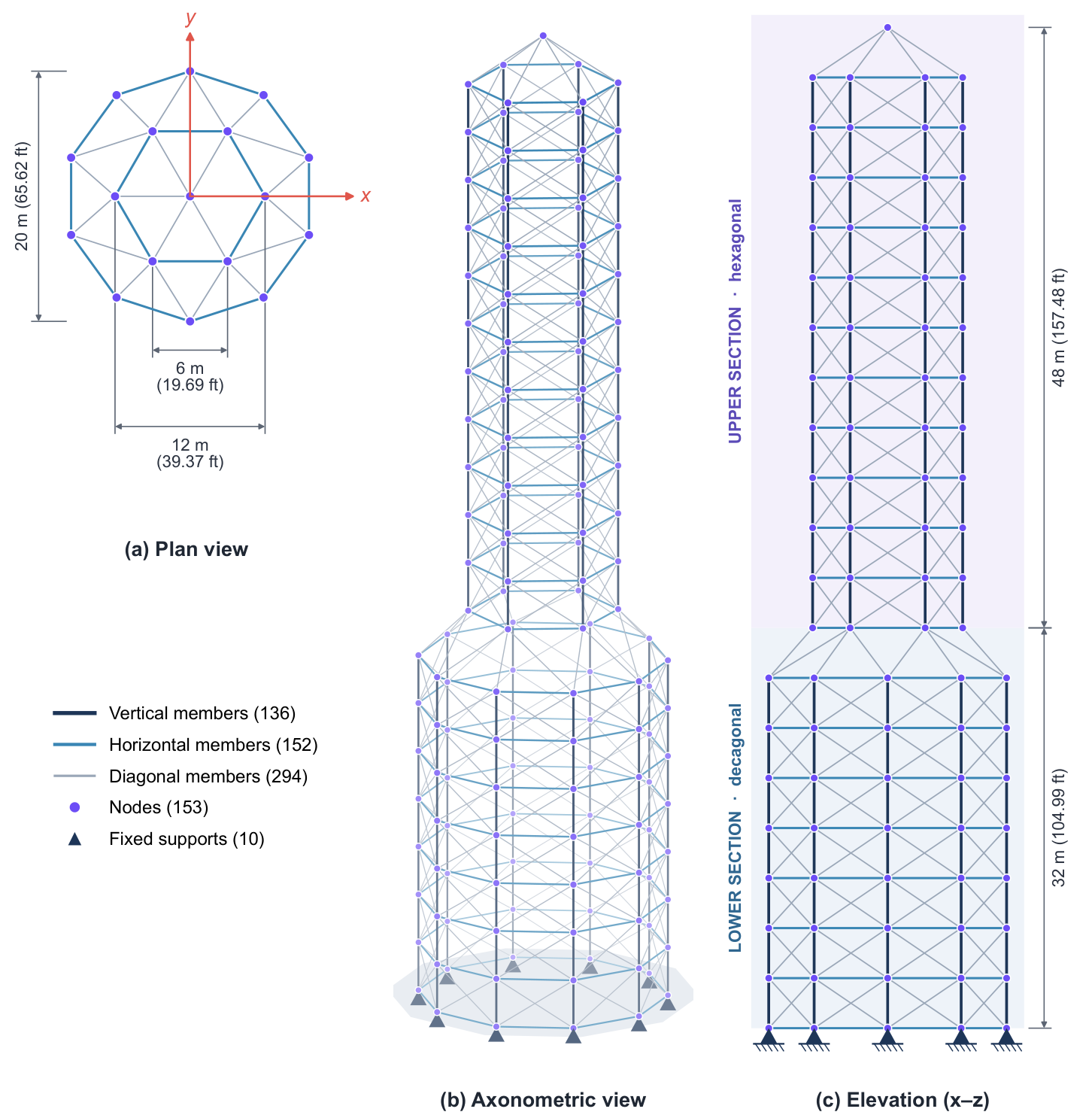}
\caption{Steel truss tower geometry from the supplied model illustration: plan, axonometric and elevation views, member groups and fixed base nodes. Dimensions and connectivity retain the original model definition.}
\end{figure}

Member-group area factors are sampled independently within ±30\% of nominal. Attached masses are obtained from the randomized gravity-load data. The input vector stores 32 area factors and 153 mass coordinates, although supported-node coordinates need not represent independent random variables. Both stiffness and mass change across the population. The material parameters reported for this model are E = 200 GPa and density 7,850 kg m⁻³.

The archived experiment retains 20 frequency-ordered modes and observes five X-direction responses under X-direction excitation in a 0--7 Hz band. The complete stored tower basis contains 429 modes, but those additional modes must not be silently substituted into the reference of an existing result. A wider, three-input, 45-channel tower experiment is a different task and is not merged into this evaluation. The first six mode shapes of the reference tower are shown in Figure 4.

\begin{figure}[!htbp]
\centering
\includegraphics[width=4.7877in]{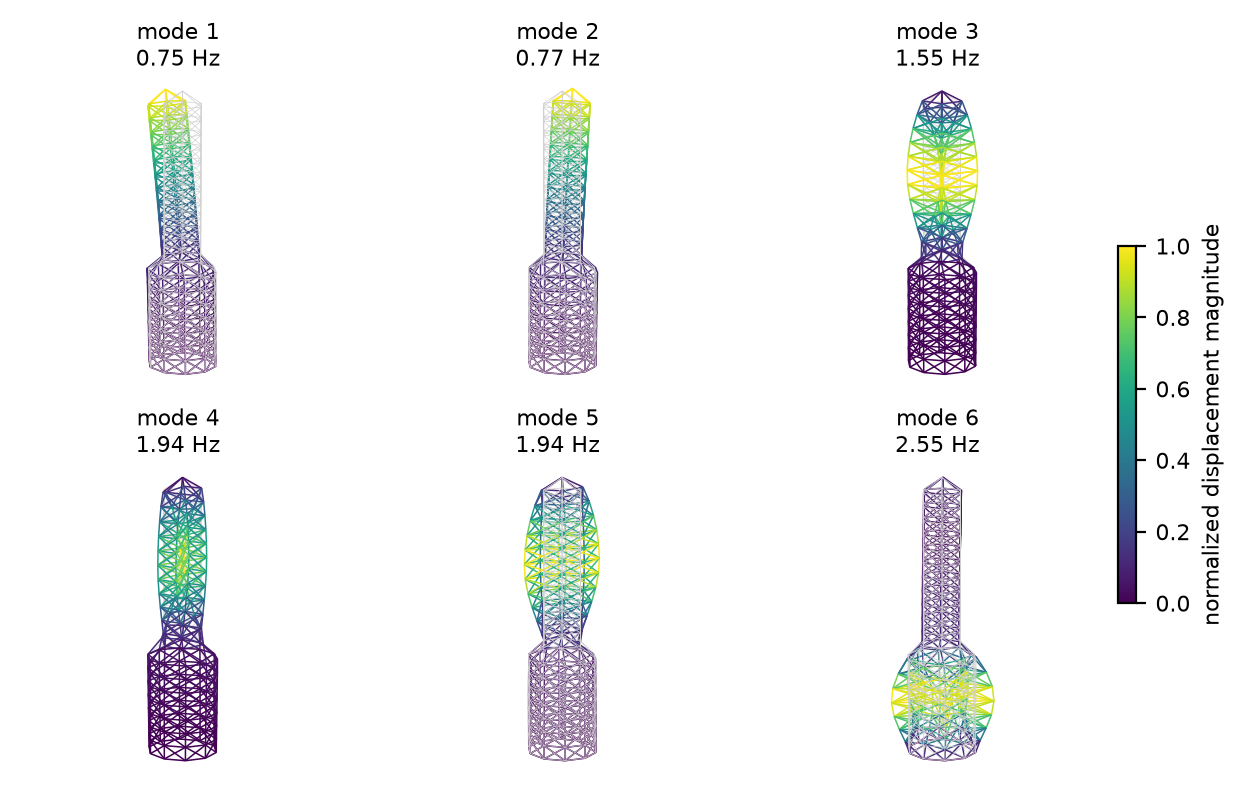}
\caption{First six mode shapes of the tower reference design, restored from the source study. Color denotes normalized displacement magnitude; the undeformed geometry is shown in gray. The two lowest modes form a closely spaced bending pair.}
\end{figure}

\subsection{7.2. Cantilevered tapered box beam}\label{cantilevered-tapered-box-beam}

The beam is 4 m long, with a square hollow section tapering from 400 mm at the clamped end to 200 mm at the free end. Wall thickness is 20 mm. The source model uses 4,000 four-node shell elements and 4,040 nodes, with steel properties E = 198 GPa, Poisson ratio 0.28 and density 7,850 kg m⁻³. Mixed-interpolation shell formulations provide the finite-element basis for this discretization {[}38, 39{]}. The tapered beam geometry is shown in Figure 5.

\begin{figure}[!htbp]
\centering
\includegraphics[width=5.3440in]{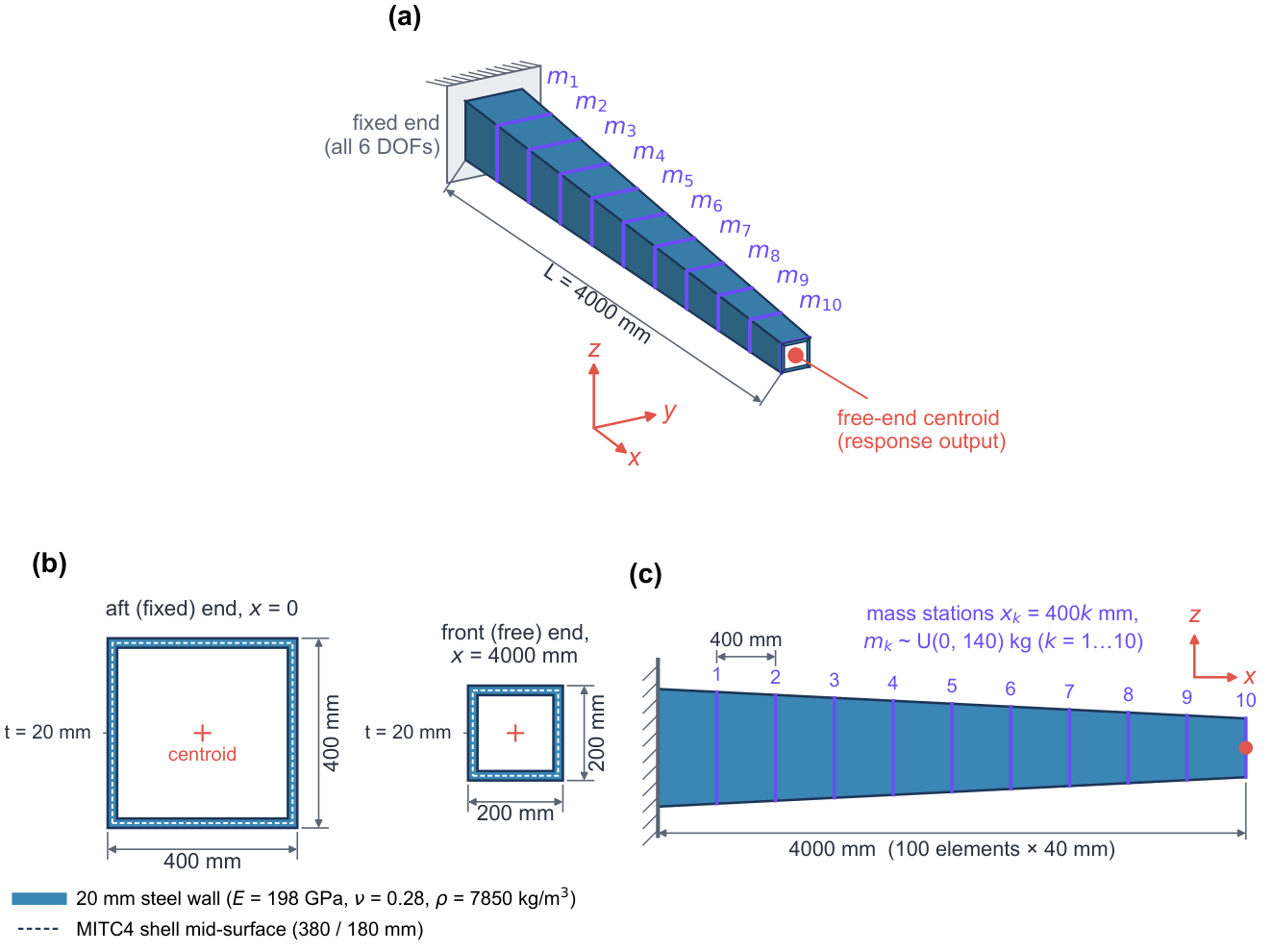}
\caption{Tapered cantilevered box-beam geometry, ten variable mass stations, end sections and free-end centroid. The source geometry and dimensions are unchanged.}
\end{figure}

Ten nonstructural masses are placed at 0.4 m station intervals, each uniformly distributed between 0 and 140 kg and spread over its cross-section nodes. The expected total added mass is 700 kg, close to the reported structural mass of 703.6 kg. Geometry, material and stiffness are fixed; only the mass distribution changes. This is a much lower-dimensional design map than the tower despite the larger finite-element (FE) model.

Twenty modes are computed with a shift-invert eigensolver {[}40, 41{]}. Nine network modes are selected by participation family rather than simply taking the nine lowest sorted eigenvalues. The reference uses saved modal information and the stated static treatment. It is not an independently certified full-shell transient solution. Modes outside the 0--100 Hz query band can still affect the low-frequency remainder. The shell discretization and representative nominal mode shapes are shown in Figures 6 and 7, respectively.

For a zero-shift solve with fixed stiffness, a factorization of K may be reused across designs. A nonzero shift requires factoring K − σM; changing M then generally changes the shifted matrix. The possibility of reuse follows from the operator and shift actually used, not from unchanged mesh connectivity alone.

\begin{figure}[!htbp]
\centering
\includegraphics[width=6.4025in]{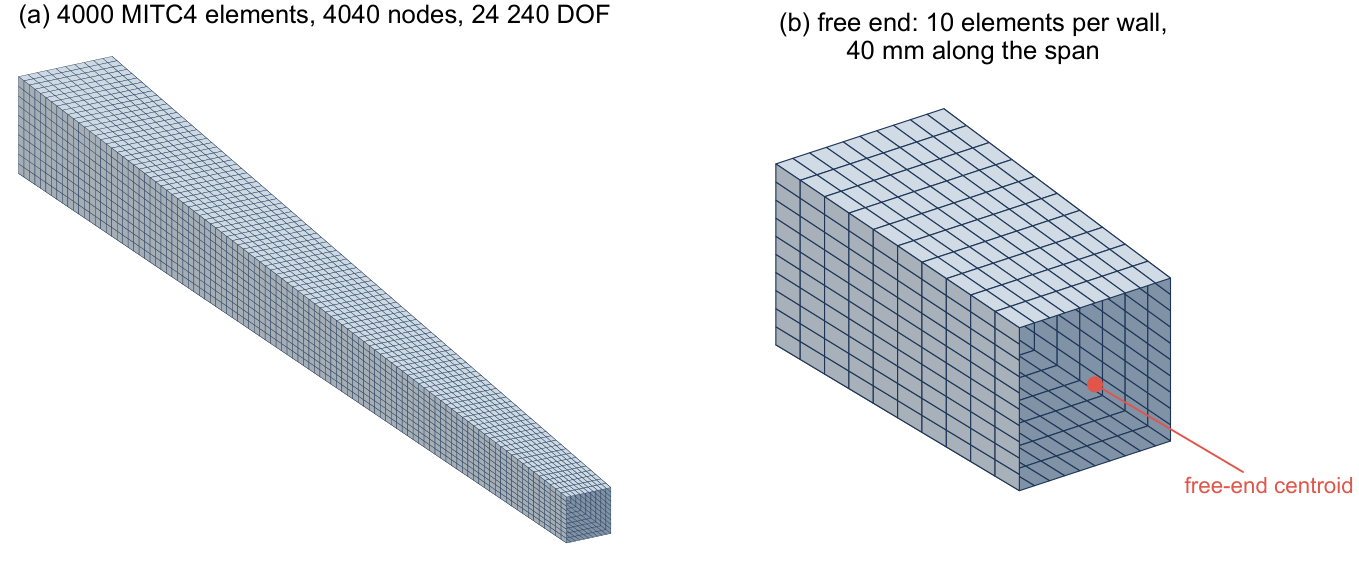}
\caption{Shell discretization of the beam, restored from the source study: (a) complete 4,000-element model and (b) free-end mesh detail. The source annotation of 24,240 DOFs is the reported total nodal count; the constrained count is distinct.}
\end{figure}

\begin{figure}[!htbp]
\centering
\includegraphics[width=6.1760in]{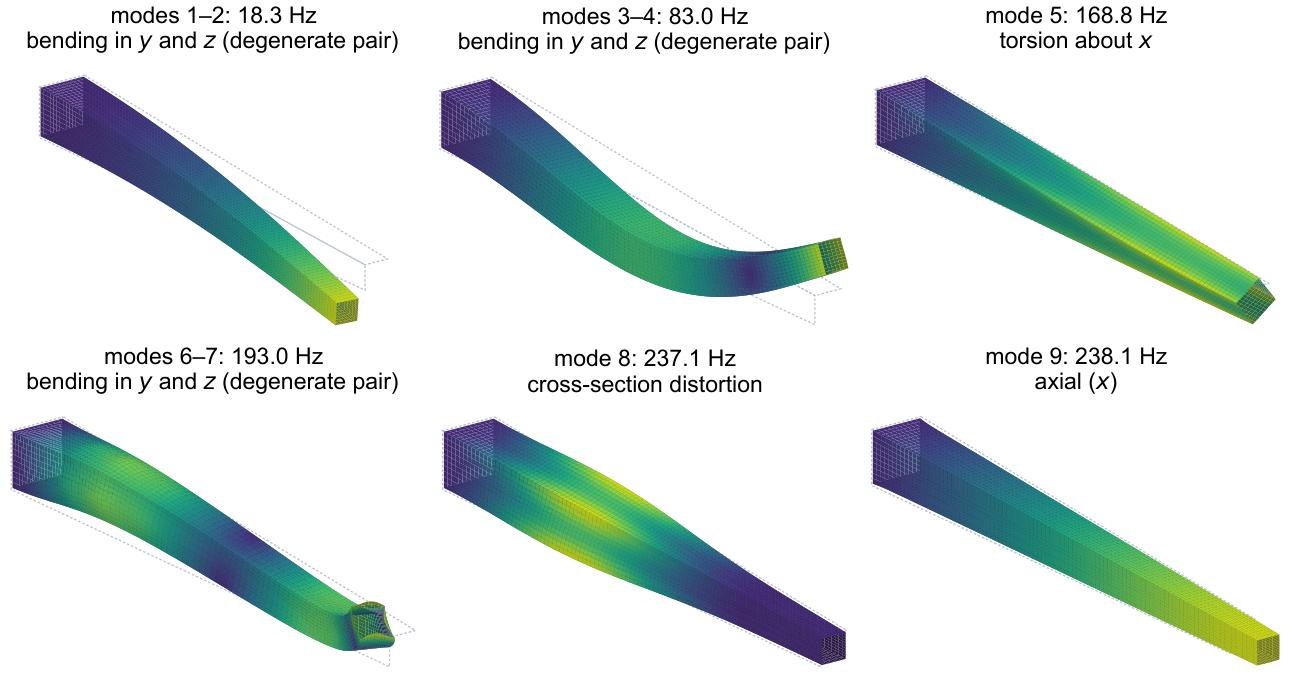}
\caption{Lowest distinct beam mode shapes for the nominal state with 70 kg at each mass station. Colors show normalized displacement magnitude. Transverse bending pairs reflect the section symmetry; the displayed mode indices describe this nominal FE solution, not a claim that all these modes lie inside the surrogate query band.}
\end{figure}

\subsection{7.3. Modal correspondence and benchmark excitation}\label{modal-correspondence-and-benchmark-excitation}

The square beam section admits repeated transverse bending modes when geometry, mass and boundary conditions preserve the relevant symmetry. An eigensolver may return different orthonormal bases of the same eigenspace. For an exactly repeated frequency and common modal damping, the sum of the associated FRF contributions is invariant under an orthogonal rotation of that subspace. Individual modal numerators need not be invariant. The source preprocessing aligns the participation directions to reduce this arbitrary labeling; modal assurance measures are useful diagnostics, not a proof of unique mode identity {[}42{]}.

\begin{equation}
\mathbf{\Phi}_{c} \mapsto \mathbf{\Phi}_{c}\mathbf{Q},\quad\mathbf{Q}^{T}\mathbf{Q} = \mathbf{I}\quad \Rightarrow \quad\left( \mathbf{L}\mathbf{\Phi}_{c} \right)\mathbf{\Phi}_{c}^{T}\mathbf{B}\ \text{is unchanged}.
\end{equation}

\emph{where} \(\mathbf{\Phi}_{\text{c}}\) contains the modes of repeated cluster c; \(\mathbf{Q}\) is an orthogonal basis rotation; \(\mathbf{I}\) is the identity matrix; \(\mathbf{L}\) is the response-selection operator; \(\mathbf{B}\) is the input/base-loading matrix.

Equation (22) gives the numerator invariance when the cluster has a common denominator. Rotating nearly repeated modes while retaining distinct diagonal frequencies does not have the same exact property. The recorded beam check reports a relative FRF change no larger than 10⁻⁵ after the chosen preprocessing; that is a reported verification value, not an independent proof for every possible clustered-mode treatment.

\subsection{Benchmark signals, not earthquake-resistant design}\label{benchmark-signals-not-earthquake-resistant-design}

The acceleration histories serve only as common, nonstationary test inputs for reconstructing response from the predicted FRF. Their historical origin in ground-motion data does not turn the cantilever into a building or make the tower a code-designed structure. No site hazard, ductility demand, collapse probability, detailing rule or earthquake-resistance decision is evaluated. The scientific question is whether the same learned structural transfer relation reproduces the numerical reference under prescribed input signals.

The benchmark acceleration histories were selected from the Pacific Earthquake Engineering Research Center (PEER) Next Generation Attenuation-West2 (NGA-West2) database using the ranked-particle optimization procedure of Behnam and Nasrollahi {[}44{]}. The seven records identified in the selection manifest reproduce the recording set used in the archived tower evaluation. For this evaluation, the selected horizontal component is applied along the global X-direction; the corresponding arrays in the X-only and three-component archives agree exactly, and their sampling intervals, component identifiers, applied multipliers, and achieved peak accelerations are consistent with the manifest. The three-component archive additionally contains the companion horizontal and vertical components, scaled using a common raw-record multiplier for each recording. This archive is not substituted for the inputs used in the archived beam evaluation, which followed its original component-specific scaling convention. The available numerical archives establish consistency of the stored signals and scaling metadata, but do not independently recover the original PEER acceleration time-series (AT2) headers or absolute timing and preprocessing records. The selected histories are used solely as reproducible, nonstationary benchmark excitations for response reconstruction; matching their suite-mean spectrum to the prescribed target does not imply that either structural model was designed or assessed for the associated seismic hazard.

The tower benchmark applies the selected horizontal component along global X. The supplied three-component archive contains that same X history, the other horizontal component and the vertical component, all multiplied by one common raw-record factor per recording. This archive must not be confused with the beam-specific convention described in the original manuscript: separate horizontal/vertical scaling and a raw-factor cap of four. The beam results and their original scaling are retained rather than silently reassigned to the newly inspected archive. The waveform metadata in Table 4 directly document the tower suite; the component/scaling association for the archived beam calculation remains part of the reproducibility record.

These inputs do not necessarily excite every part of the declared FRF band strongly. A sampled record with interval Δt has Nyquist frequency 1/(2Δt); interpolation cannot establish measured information above its original usable bandwidth. Consequently, low response-history error under a particular suite is not uniform FRF validation across 0--100 Hz. That limitation is especially important for the beam and is separate from its lack of a seismic-design interpretation.

\subsection{Candidate selection and scaling}\label{candidate-selection-and-scaling}

The recorded screening retained 29 horizontal components from 18 recordings with 180 ≤ Vₛ₃₀ \textless{} 360 m s⁻¹ and raw peak ground acceleration (PGA) of at least 0.10g. A seven-record suite was chosen using 100 restarts of 1,000 ranked-particle iterations, seed 20260712, and optimizer factors sᵢ in {[}0.5, 2.0{]}. Two components with the same record sequence number were treated as duplicates. The saved catalog contains 5\%-damped spectra at 201 oscillator periods from 0 to 4 s. The selection procedure uses the Eurocode-8-shaped ground-type-C target recorded in the archive; it is not fitted to a beam design requirement {[}44{]}.

For a raw history with \(P\), the applied multiplier is \(z_{\text{q,o}}\) = \(s_{\text{o}}a_{\text{d}}\)/\(P\), where \(a_{\text{d}}\) = 0.35g. Thus aᵢ,scaled(t) = \(z_{\text{q,o}}\)aᵢ,raw(t), and the achieved X-component PGA is \(s_{\text{o}}a_{\text{d}}\). The bound on \(s_{\text{o}}\) is not a bound on \(z_{\text{q,o}}\): the largest raw multiplier in Table 4 is 5.473. The already-scaled arrays must not be multiplied by \(s_{\text{o}}\) a second time. Selection favors a suite-mean spectrum near the target and penalizes duplicate recordings and spectral constraint violations; the archived objective is 1.539408.

Reconstruction from the saved spectrum catalog gives suite-mean/target ratios of 0.9000--1.0992 at the nonzero tabulated periods. At zero period the ratio is 1.1907, so the upper ±10\% band is not met there. The mean achieved X PGA is 0.4793g. These values reproduce the recorded selection checks and describe this benchmark only. No blanket spectrum-compliance or structural-safety claim follows from them. Figure 8 shows the corresponding spectrum-based record selection.

\subsection{7.4. Record selection, scaling and numerical reference}\label{record-selection-scaling-and-numerical-reference}

\Needspace{7\baselineskip}
\begin{table}[H]
\tablecaption{Seven NGA-West2 benchmark recordings; RSN denotes record sequence number. The records are verified against the supplied selection manifest and X-component archive. Event names identify signal provenance, not a seismic-design case. q is the multiplier applied to the raw X accelerogram; s is the optimizer scale factor. Peak acceleration is measured before the structural-band taper.}

\centering
\begin{tabular}{@{}
  >{\raggedright\arraybackslash}p{(\columnwidth - 12\tabcolsep) * \real{0.1710}}
  >{\raggedright\arraybackslash}p{(\columnwidth - 12\tabcolsep) * \real{0.3530}}
  >{\raggedright\arraybackslash}p{(\columnwidth - 12\tabcolsep) * \real{0.0799}}
  >{\raggedright\arraybackslash}p{(\columnwidth - 12\tabcolsep) * \real{0.1118}}
  >{\raggedright\arraybackslash}p{(\columnwidth - 12\tabcolsep) * \real{0.0799}}
  >{\raggedright\arraybackslash}p{(\columnwidth - 12\tabcolsep) * \real{0.0991}}
  >{\raggedright\arraybackslash}p{(\columnwidth - 12\tabcolsep) * \real{0.1054}}@{}}
\toprule\noalign{}
\begin{minipage}[b]{\linewidth}\raggedright
\textbf{RSN / X component}
\end{minipage} & \begin{minipage}[b]{\linewidth}\raggedright
\textbf{Event, year; station}
\end{minipage} & \begin{minipage}[b]{\linewidth}\raggedright
\textbf{Δt\\
(s)}\strut
\end{minipage} & \begin{minipage}[b]{\linewidth}\raggedright
\textbf{Length¹\\
(s)}\strut
\end{minipage} & \begin{minipage}[b]{\linewidth}\raggedright
\textbf{s}
\end{minipage} & \begin{minipage}[b]{\linewidth}\raggedright
\textbf{q}
\end{minipage} & \begin{minipage}[b]{\linewidth}\raggedright
\textbf{PGAₓ\\
(g)}\strut
\end{minipage} \\
\midrule\noalign{}
6 / I-ELC270 & \begin{minipage}[t]{\linewidth}\raggedright
Imperial Valley-02, 1940\\
El Centro Array \#9\strut
\end{minipage} & 0.010 & 53.460 & 2.000 & 3.322 & 0.700 \\
20 / H-FRN044 & \begin{minipage}[t]{\linewidth}\raggedright
Northern Calif-03, 1954\\
Ferndale City Hall\strut
\end{minipage} & 0.005 & 40.000 & 1.292 & 2.768 & 0.452 \\
30 / C05355 & \begin{minipage}[t]{\linewidth}\raggedright
Parkfield, 1966\\
Cholame - Shandon Array \#5\strut
\end{minipage} & 0.010 & 44.000 & 0.500 & 0.475 & 0.175 \\
34 / C-FRN314 & \begin{minipage}[t]{\linewidth}\raggedright
Northern Calif-05, 1967\\
Ferndale City Hall\strut
\end{minipage} & 0.005 & 92.975 & 1.300 & 4.204 & 0.455 \\
95 / A-ESO090 & \begin{minipage}[t]{\linewidth}\raggedright
Managua Nicaragua-01, 1972\\
Managua ESSO\strut
\end{minipage} & 0.005 & 45.690 & 1.224 & 1.152 & 0.428 \\
97 / PHN180 & \begin{minipage}[t]{\linewidth}\raggedright
Point Mugu, 1973\\
Port Hueneme\strut
\end{minipage} & 0.005 & 23.185 & 1.998 & 5.473 & 0.699 \\
100 / A-SJB123 & \begin{minipage}[t]{\linewidth}\raggedright
Hollister-03, 1974\\
San Juan Bautista 24 Polk St\strut
\end{minipage} & 0.005 & 20.755 & 1.271 & 3.830 & 0.445 \\
\bottomrule\noalign{}
\end{tabular}
\end{table}

¹ Length follows the manifest convention NΔt; the final sample is at (N − 1)Δt. The X samples in the X-only and X/Y/Z archives agree exactly. The stored PGA values agree with maxima computed from the supplied X arrays. These checks establish archive consistency, not independent authentication of the original recordings.

\begin{figure}[!htbp]
\centering
\includegraphics[width=4.8620in]{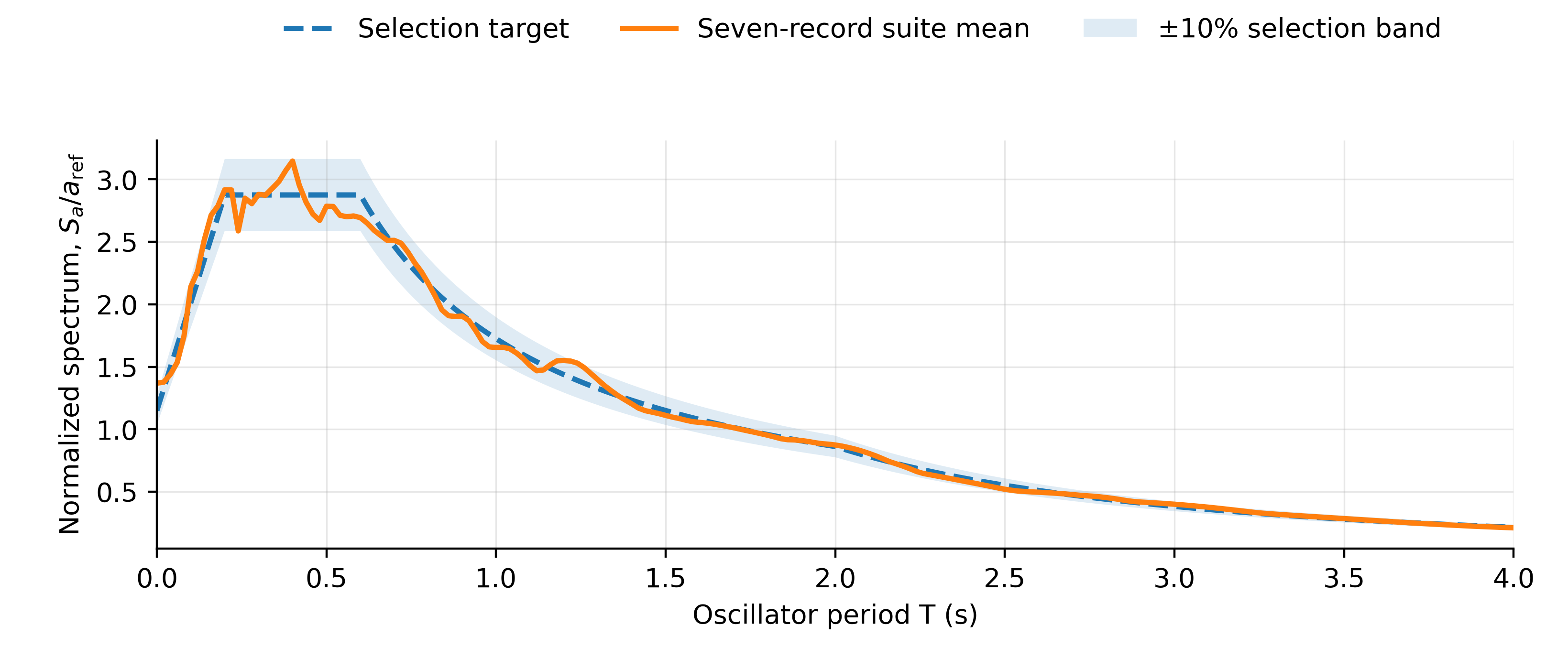}
\caption{Spectrum-based selection of the documented tower inputs. The suite mean and selection target are normalized by the reference acceleration 0.35g; the shaded region is the ±10\% selection band. The zero-period excess is retained. The plot is reconstructed from the supplied catalog and scale factors, not from new or synthesized accelerograms. The beam-specific spectrum is not substituted from this archive.}
\end{figure}

The numerical reference is the specified modal calculation with the same loading, damping, band limit and signal processing used for each surrogate. It is ``closed form'' at the modal-response level; it is not a statement that the finite-element idealization or truncated representation is an exact physical model. Reference errors and neural approximation errors must not be combined under one label.

The archived verification compared modal reconstruction with direct Newmark integration of the complete tower for two designs and seven inputs. Newmark integration is a standard time-discretization method {[}45{]}. The reported differences are retained in Table 5. They establish agreement for those cases, but attributing every difference to time integration would require a time-step and truncation convergence study. In particular, the largest reported discrepancy is not negligible relative to every small difference between surrogate methods.

\Needspace{7\baselineskip}
\begin{table}[H]
\tablecaption{Source-reported tower reference checks against direct time integration; percentages are differences between numerical solution paths, not neural prediction errors.}

\centering
\begin{tabular}{@{}
  >{\raggedright\arraybackslash}p{(\columnwidth - 4\tabcolsep) * \real{0.4445}}
  >{\raggedright\arraybackslash}p{(\columnwidth - 4\tabcolsep) * \real{0.2778}}
  >{\raggedright\arraybackslash}p{(\columnwidth - 4\tabcolsep) * \real{0.2778}}@{}}
\toprule\noalign{}
\begin{minipage}[b]{\linewidth}\raggedright
\textbf{Quantity}
\end{minipage} & \begin{minipage}[b]{\linewidth}\raggedright
\textbf{Mean difference (\%)}
\end{minipage} & \begin{minipage}[b]{\linewidth}\raggedright
\textbf{Maximum difference (\%)}
\end{minipage} \\
\midrule\noalign{}
Relative displacement & 0.016 & 0.31 \\
Relative velocity & 0.044 & 0.70 \\
Relative acceleration & 0.13 & 1.45 \\
Absolute acceleration & 0.13 & 1.54 \\
\bottomrule\noalign{}
\end{tabular}
\end{table}

Additional archived checks report that the all-mode static sum matches the direct tower static solution to approximately 10⁻¹², and that the first twenty frequencies in the rebuilt database differ from the earlier database by at most 2.3 × 10⁻⁷ relatively. These checks support specific identities and data consistency. They do not establish mesh convergence, failure-free inference across every design, or the equivalence of all historical checkpoint implementations.

For the beam, modal normalization, mode pairing and static reconstruction are part of the reference definition. The present manuscript does not claim an independently verified full-shell time-history solution. Keeping this distinction explicit is more useful than calling every saved modal reconstruction ``ground truth'' without qualification.

\FloatBarrier
\section{8. Response accuracy}\label{response-accuracy}

\subsection{8.1. Metrics and aggregation}\label{metrics-and-aggregation}

For a reference history \(y\) and prediction \(\widehat{y}\) sampled on the same retained time interval, the response error is

\begin{equation}
\epsilon_{y} = 100\frac{\parallel \widehat{\mathbf{y}} - \mathbf{y} \parallel_{2}}{\parallel \mathbf{y} \parallel_{2}},\quad\quad\epsilon_{\text{peak}} = 100\frac{\text{abs}\left( P\left( \widehat{y} \right) - P(y) \right)}{P(y)}.
\end{equation}

\emph{where} \(\epsilon_{\text{y}}\) is the relative response-history L₂ error; \(\epsilon_{\text{peak}}\) is the relative peak-amplitude error; \(y\) is the reference history; \(\widehat{y}\) is the predicted history; \(P(y)\) returns the peak absolute amplitude; \(\text{abs}\) denotes absolute value.

Here \(P\) is the peak absolute amplitude of history \(y\), and abs denotes absolute value. The histories are assessed separately for relative displacement, velocity, relative acceleration and absolute acceleration. Near-zero reference histories require an absolute-error safeguard; its threshold and any inactive-output mask belong to the evaluation specification. No new exclusion or threshold was introduced while rewriting these results. A zero denominator is not a finite percentage and should not be silently removed by an averaging function.

The following tables preserve the response-specific means of the archived evaluation. Their maximum column concerns the displacement metric, as declared by the source table caption; it is not an all-response worst-case certificate. Input records and output cases are correlated within each of the ten evaluated designs. Figure 9 summarizes the aggregate results; Figures 10--12 restore the source FRF, response-history and peak-agreement evidence, while Figure 2 restores the convergence evidence. No curves are inferred from the aggregate tables.

\Needspace{7\baselineskip}
\begin{table}[H]
\tablecaption{Tower response-history errors (\%): ten test designs, seven X-direction benchmark inputs and five response locations. Peak and maximum columns refer to relative displacement.}

\centering
\begin{tabular}{@{}
  >{\raggedright\arraybackslash}p{(\columnwidth - 12\tabcolsep) * \real{0.3175}}
  >{\raggedright\arraybackslash}p{(\columnwidth - 12\tabcolsep) * \real{0.1111}}
  >{\raggedright\arraybackslash}p{(\columnwidth - 12\tabcolsep) * \real{0.1111}}
  >{\raggedright\arraybackslash}p{(\columnwidth - 12\tabcolsep) * \real{0.1270}}
  >{\raggedright\arraybackslash}p{(\columnwidth - 12\tabcolsep) * \real{0.1270}}
  >{\raggedright\arraybackslash}p{(\columnwidth - 12\tabcolsep) * \real{0.1032}}
  >{\raggedright\arraybackslash}p{(\columnwidth - 12\tabcolsep) * \real{0.1032}}@{}}
\toprule\noalign{}
\begin{minipage}[b]{\linewidth}\raggedright
\textbf{Method}
\end{minipage} & \begin{minipage}[b]{\linewidth}\raggedright
\textbf{Disp.}
\end{minipage} & \begin{minipage}[b]{\linewidth}\raggedright
\textbf{Vel.}
\end{minipage} & \begin{minipage}[b]{\linewidth}\raggedright
\textbf{Rel. acc.}
\end{minipage} & \begin{minipage}[b]{\linewidth}\raggedright
\textbf{Abs. acc.}
\end{minipage} & \begin{minipage}[b]{\linewidth}\raggedright
\textbf{Peak}
\end{minipage} & \begin{minipage}[b]{\linewidth}\raggedright
\textbf{Max}
\end{minipage} \\
\midrule\noalign{}
ReLU* & 32.29 & 27.66 & 21.55 & 22.67 & 4.65 & 144.0 \\
GELU* & 13.47 & 11.87 & 10.23 & 10.54 & 2.26 & 28.6 \\
SIREN & 4.54 & 5.54 & 7.82 & 8.17 & 2.36 & 13.8 \\
\textbf{Anchored SIREN} & \textbf{0.72} & \textbf{0.91} & \textbf{1.27} & 1.34 & \textbf{0.60} & \textbf{2.1} \\
Modal Rational & 0.73 & 1.01 & 1.38 & 1.33 & 0.52 & 5.3 \\
PGMON & 0.74 & 0.95 & 1.28 & 1.26 & 0.58 & 3.8 \\
MOSAIC-MR & 0.76 & 1.04 & 1.37 & 1.30 & 0.50 & 5.4 \\
MOSAIC-PGMON & 0.73 & 0.92 & 1.28 & \textbf{1.27} & \textbf{0.60} & 3.6 \\
\bottomrule\noalign{}
\end{tabular}
\end{table}

\emph{* Historical labels and values are retained; they must not be reassigned to the nodal-modal output task in Eq. (1). Section 3.2 states the unresolved implementation association. The MR-family rows likewise describe the archived conditional-frequency evaluation, not the later training campaign.}

The modal configurations have similar mean tower displacement errors: 0.73--0.76\%. Their absolute-acceleration means lie between 1.26\% and 1.33\%. The anchored SIREN gives the smallest listed mean displacement error, 0.72\%, and the smallest listed displacement maximum, 2.1\%. PGMON and MOSAIC-PGMON improve some acceleration or maximum-displacement entries relative to the archived MR row, but no method wins every metric. Differences of a few hundredths of a percentage point should not be presented as statistically established rankings from one seed.

\subsection{8.2. Tower comparison}\label{tower-comparison}

The figure separates two observations. First, the archived spectral comparators labeled ReLU, GELU and SIREN have much larger errors than the modal-aware configurations. Second, the differences within the modal-aware group are relatively small and response dependent. The former observation concerns complete implementations with different information and supervision; it is not an isolated estimate of the benefit of an analytical decoder or of the author-specified nodal-modal baseline.

The acceleration means are larger than the corresponding displacement means for the modal configurations. For example, Modal Rational reports 0.73\% mean displacement error and 1.38\% relative-acceleration error. PGMON changes these to 0.74\% and 1.28\%. This trade-off is compatible with the frequency weighting of derivative response, but the table alone does not identify which modes or channels caused the change. Section 9 supplies an error analysis that makes those alternatives explicit.

The displacement maxima are several times the means. That observation justifies keeping the difficult cases visible, not declaring the entire population unreliable or the worst case controlled. The archived minimum correlations, rounded to 0.999--1.000 for the modal-aware models, describe waveform similarity; correlation near one does not guarantee a small amplitude or peak error.

For this restricted tower task, modal information and analytical reconstruction support low average response errors. The additional corrections change particular metrics without uniform dominance. This comparison does not require every extension to improve, and the observed maxima are retained rather than replaced by a more favorable checkpoint.

\subsection{8.3. Beam response accuracy}\label{beam-response-accuracy}

\Needspace{7\baselineskip}
\begin{table}[H]
\tablecaption{Beam response-history errors (\%): ten test designs, seven three-component benchmark inputs and the free-end translational response. Peak and maximum columns retain the source displacement definition.}

\centering
\begin{tabular}{@{}
  >{\raggedright\arraybackslash}p{(\columnwidth - 12\tabcolsep) * \real{0.3175}}
  >{\raggedright\arraybackslash}p{(\columnwidth - 12\tabcolsep) * \real{0.1111}}
  >{\raggedright\arraybackslash}p{(\columnwidth - 12\tabcolsep) * \real{0.1111}}
  >{\raggedright\arraybackslash}p{(\columnwidth - 12\tabcolsep) * \real{0.1270}}
  >{\raggedright\arraybackslash}p{(\columnwidth - 12\tabcolsep) * \real{0.1270}}
  >{\raggedright\arraybackslash}p{(\columnwidth - 12\tabcolsep) * \real{0.1032}}
  >{\raggedright\arraybackslash}p{(\columnwidth - 12\tabcolsep) * \real{0.1032}}@{}}
\toprule\noalign{}
\begin{minipage}[b]{\linewidth}\raggedright
\textbf{Method}
\end{minipage} & \begin{minipage}[b]{\linewidth}\raggedright
\textbf{Disp.}
\end{minipage} & \begin{minipage}[b]{\linewidth}\raggedright
\textbf{Vel.}
\end{minipage} & \begin{minipage}[b]{\linewidth}\raggedright
\textbf{Rel. acc.}
\end{minipage} & \begin{minipage}[b]{\linewidth}\raggedright
\textbf{Abs. acc.}
\end{minipage} & \begin{minipage}[b]{\linewidth}\raggedright
\textbf{Peak}
\end{minipage} & \begin{minipage}[b]{\linewidth}\raggedright
\textbf{Max}
\end{minipage} \\
\midrule\noalign{}
ReLU* & 0.365 & 0.498 & 0.604 & 0.236 & 0.103 & 1.07 \\
GELU* & 0.281 & 0.353 & 0.376 & 0.098 & 0.035 & 1.18 \\
SIREN & 3.959 & 5.716 & 6.016 & 2.076 & 0.502 & 13.70 \\
Anchored SIREN & 0.229 & 0.247 & 0.268 & 0.056 & 0.024 & 0.56 \\
Modal Rational & 0.116 & 0.113 & 0.109 & 0.011 & 0.009 & 0.67 \\
PGMON & 0.034 & 0.033 & 0.032 & 0.010 & 0.008 & 0.12 \\
MOSAIC-MR & 0.116 & 0.112 & 0.107 & 0.009 & 0.007 & 0.68 \\
\textbf{MOSAIC-PGMON} & \textbf{0.029} & \textbf{0.028} & \textbf{0.026} & \textbf{0.008} & \textbf{0.007} & \textbf{0.10} \\
\bottomrule\noalign{}
\end{tabular}
\end{table}

\emph{* Same implementation-provenance qualification as Table 6. Values are retained exactly at their supplied precision. The beam's inactive-output convention and richer training recipe prevent treating it as a matched replication of the tower.}

The recorded beam results favor the PGMON configurations more clearly than the tower results do. Relative to the archived MR displacement mean of 0.116\%, PGMON gives 0.034\% and MOSAIC-PGMON 0.029\%. Relative-acceleration means change from 0.109\% to 0.032\% and 0.026\%. These are approximately 71--76\% reductions in the listed mean errors, but they are configuration-level comparisons, not isolated estimates of the effect of one correction head.

\begin{figure}[!htbp]
\centering
\includegraphics[width=5.5071in]{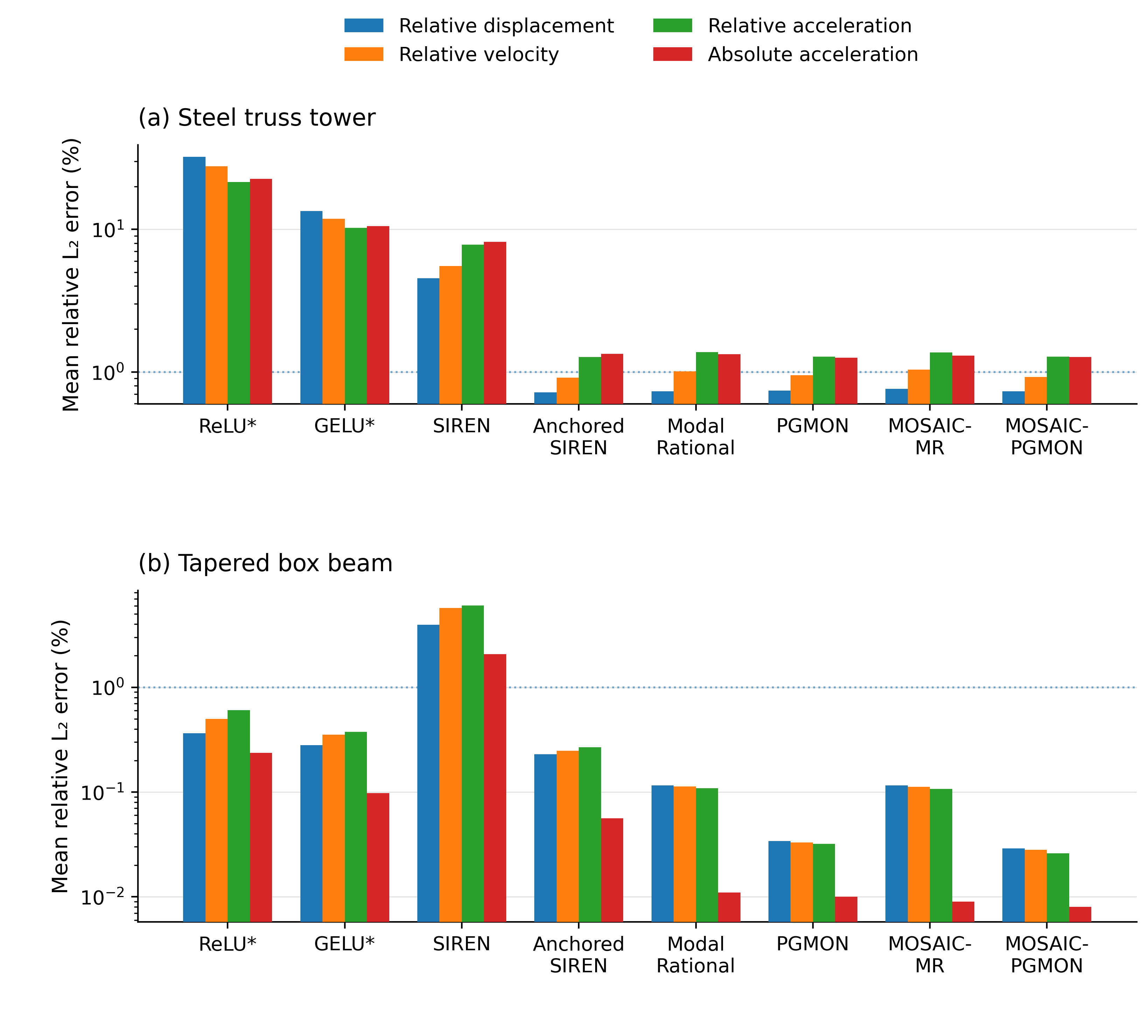}
\caption{Mean relative response-history L₂ errors for (a) the tower and (b) the beam, redrawn from Tables 6 and 7. Each bar represents one response quantity; axes are logarithmic and the dotted line marks 1\%, not an acceptance threshold. The legend is outside the data areas. Asterisks retain the implementation qualification attached to the ReLU/GELU rows. The numerical entries are unchanged from the source comparison.}
\end{figure}

MOSAIC-MR is close to MR: its displacement mean remains 0.116\%, while its relative-acceleration mean changes from 0.109\% to 0.107\%. The results therefore do not support a universal benefit from a bounded spectral discrepancy. They show that its additional effect depends on the parent and the task. The reported PGMON improvements also coexist with different training and internal parameterization choices; those dependencies remain part of the finding.

The low absolute-acceleration errors need careful interpretation. When the reference and prediction use the same direct input contribution, their absolute- and relative-acceleration error numerators are identical. Their percentage denominators differ. Thus, 0.008\% absolute-acceleration error does not establish 0.008\% accuracy of every modal quantity, every relative-acceleration channel or every FRF query.

Nor does this example establish that the modal models become more advantageous simply because the band extends to 100 Hz. The beam has ten varying masses, a fixed stiffness, aligned participation families and a different training objective. A bandwidth-only claim would require a within-system comparison in which those other choices are held fixed. What the present evidence shows is that a large FE discretization can coexist with an accurately learned, comparatively small design-to-modal map.

The reported beam minima of the correlation coefficient round to one for several modal configurations. The response-error and peak-error columns remain necessary because high correlation can coexist with scaling error. As with the tower, neither rounding nor ten sampled designs supplies a uniform engineering tolerance over the full population.

\subsection{8.4. Beam--tower contrast}\label{beamtower-contrast}

The contrast between the two panels of Figure 9 is useful because the structures challenge different parts of the learning problem. The tower varies stiffness and mass in a high-dimensional input space; the beam varies only ten masses and uses a participation-based modal organization. Conversely, the beam has nine transfer channels and a wider query band. The FE degree-of-freedom count alone therefore does not rank their learning difficulty.

For the archived MR rows, tower and beam displacement means are 0.73\% and 0.116\%, respectively; their relative-acceleration means are 1.38\% and 0.109\%. This difference is real in the supplied tables, but its causal attribution is not settled by those two numbers. A larger input dimension, mode-label changes, differing output magnitudes, omitted dynamics and different optimization recipes can all contribute. They are distinguishable mechanisms, not interchangeable explanations.

A further asymmetry is the beam's handling of weak channels. The training appendix reports six of nine transfer entries below the inactive-output variability threshold. These entries are not six failed structural responses; they are small transfer channels under that normalization rule. Nevertheless, masking changes what a pooled score emphasizes. The mask, its training-only construction and the inclusion of all physical components in time-history reconstruction must remain explicit.

The most defensible empirical statement is that analytical modal structure is effective in the declared experiments, while additional corrections are more useful in some configurations than others. The evidence does not establish superiority over every well-tuned generic method, nor the impossibility of learning the tower accurately. It motivates the error decomposition in the next section and a later matched study of operator implementations.

\subsection{8.5. Frequency-response localization and reconstructed histories}\label{frequency-response-localization-and-reconstructed-histories}

Figure 10 restores the source FRF comparison at the tower apex for design 1944 and at the beam free-end centroid for design 7. The top panels show the response magnitude; the lower panels normalize the absolute complex-FRF discrepancy by the maximum magnitude of that same reference FRF. This normalization is not a pointwise relative error and is not the response-history metric of Tables 6 and 7. The examples show where the discrepancy occurs without replacing the multi-design evaluation.

\begin{figure}[!htbp]
\centering
\includegraphics[width=5.1028in]{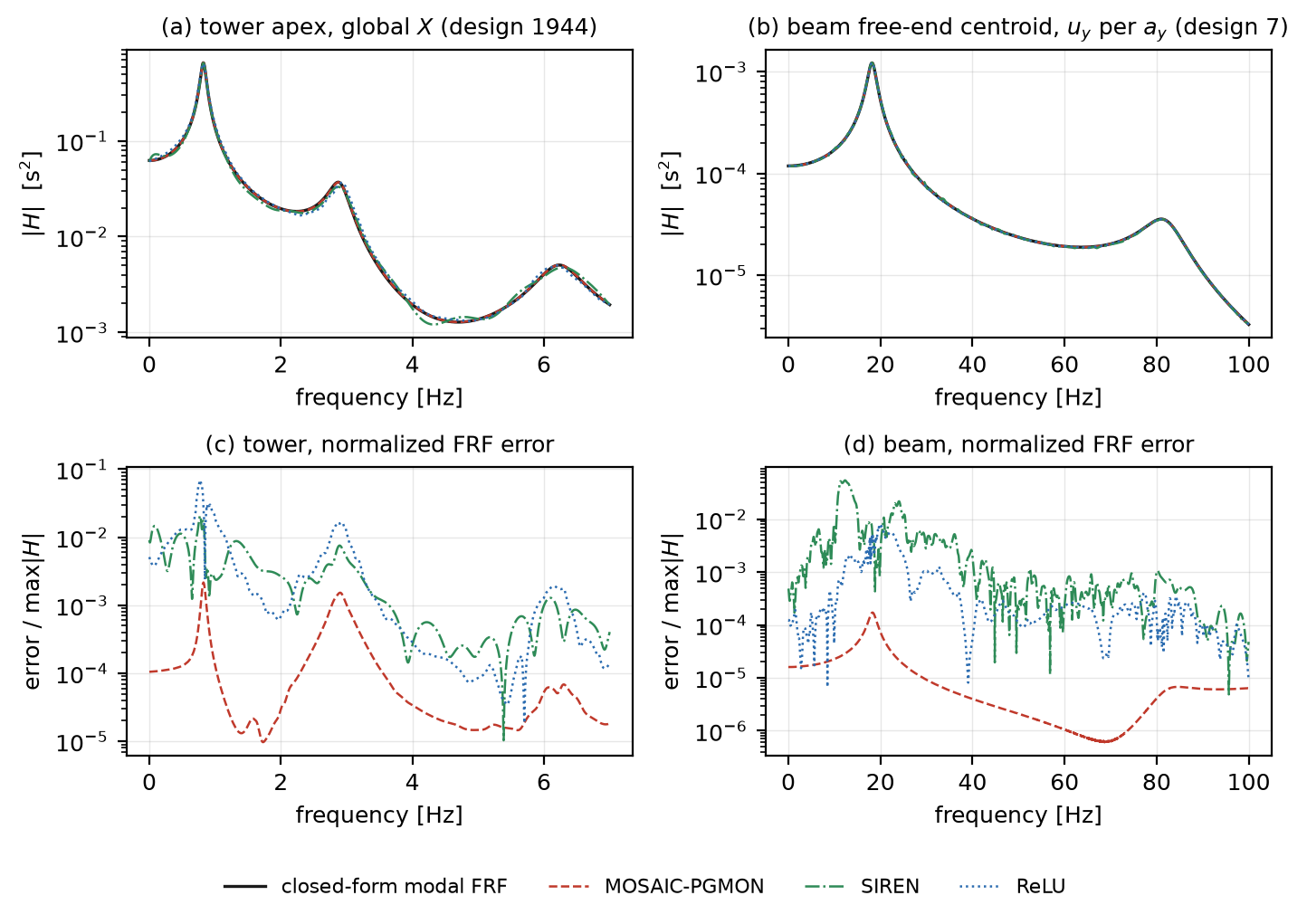}
\caption{Source FRF comparison, restored without changing its curves: (a, b) magnitudes for one held-out tower and beam design; (c, d) absolute complex-FRF error divided by the peak magnitude of the corresponding reference FRF. Logarithmic ordinate scales and an external legend are retained. These are representative examples, not distributions over all designs.}
\end{figure}

Peak locations, local phase effects, and numerator errors can all influence the discrepancy near a resonance. A visually aligned magnitude peak does not establish an accurate complex response, while a localized error peak does not uniquely identify a frequency-prediction fault. In particular, the conditional-frequency configurations described earlier must not be assigned an anchor-learning error that their inference path does not contain. The source curves are therefore interpreted as response evidence, not a standalone causal diagnosis.

Figure 11 restores the corresponding time-history evidence under the input labeled RSN6. It displays short response windows and their signed prediction errors rather than the full records. The tower and beam use different ordinate scales and native units, which are retained explicitly. Agreement in one window supports the visual comparison; the complete response metrics remain those in the tables.

\begin{figure}[!htbp]
\centering
\includegraphics[width=4.6929in]{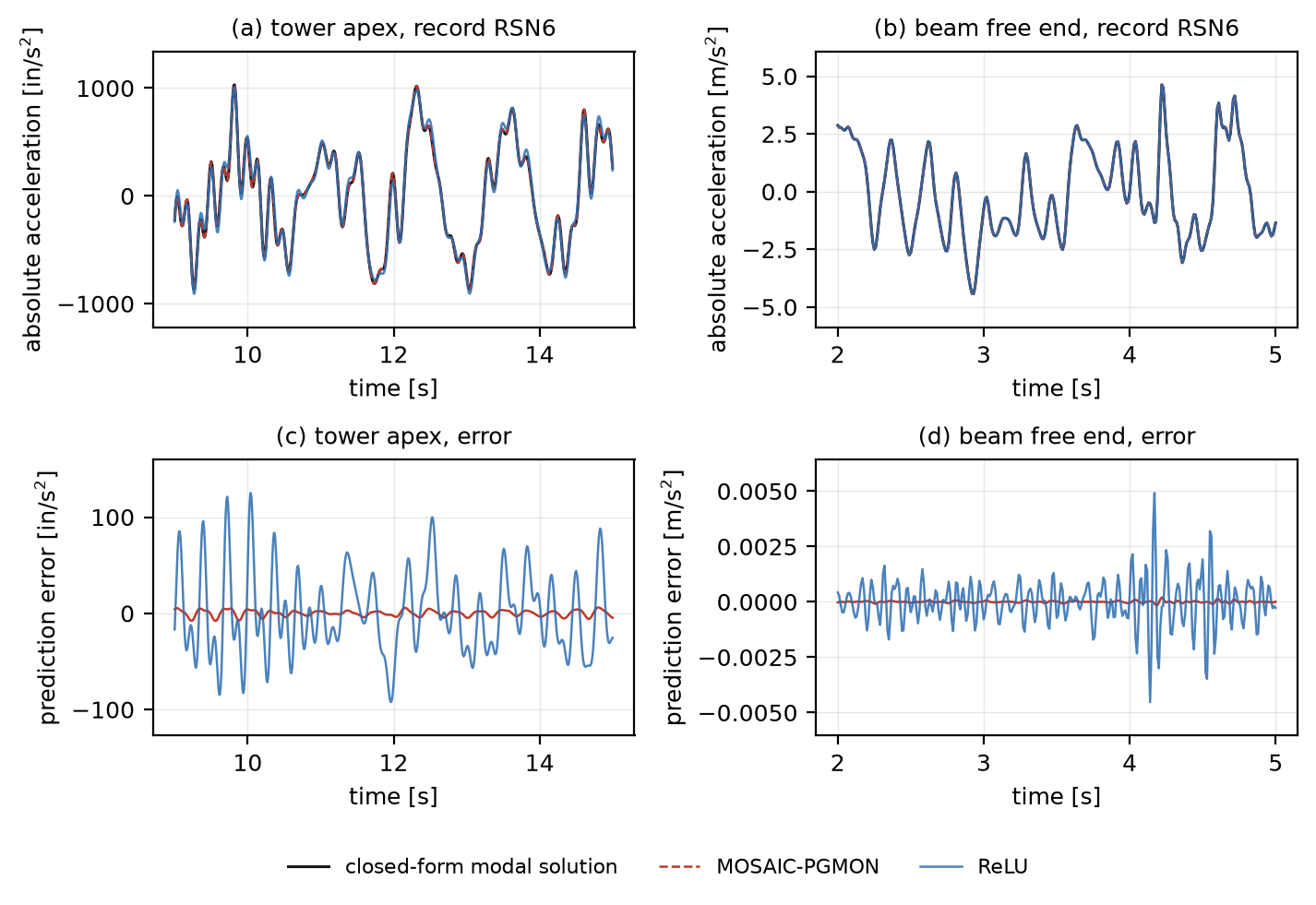}
\caption{Representative absolute-acceleration responses under the archived RSN6 input: (a) tower apex (in/s²); (b) beam free-end response (m/s²); (c, d) signed prediction errors in the corresponding units. The plotted time windows, predictions and modal-reference traces are retained from the source. No waveform synthesis or unit rescaling was performed.}
\end{figure}

\Needspace{0.55\textheight}
\subsection{8.6. Peak-response agreement}\label{peak-response-agreement}

The peak comparison in Figure 12 complements both time-history error and correlation. It retains the source points for the four displayed methods across the evaluated designs, inputs, and response locations. Points near the identity indicate agreement in peak absolute acceleration, but this statistic does not constrain phase or the rest of the time history. The figure concerns acceleration peaks; the peak column in Tables 6 and 7 retains its original displacement definition.

\begin{figure}[H]
\centering
\includegraphics[width=5.5076in]{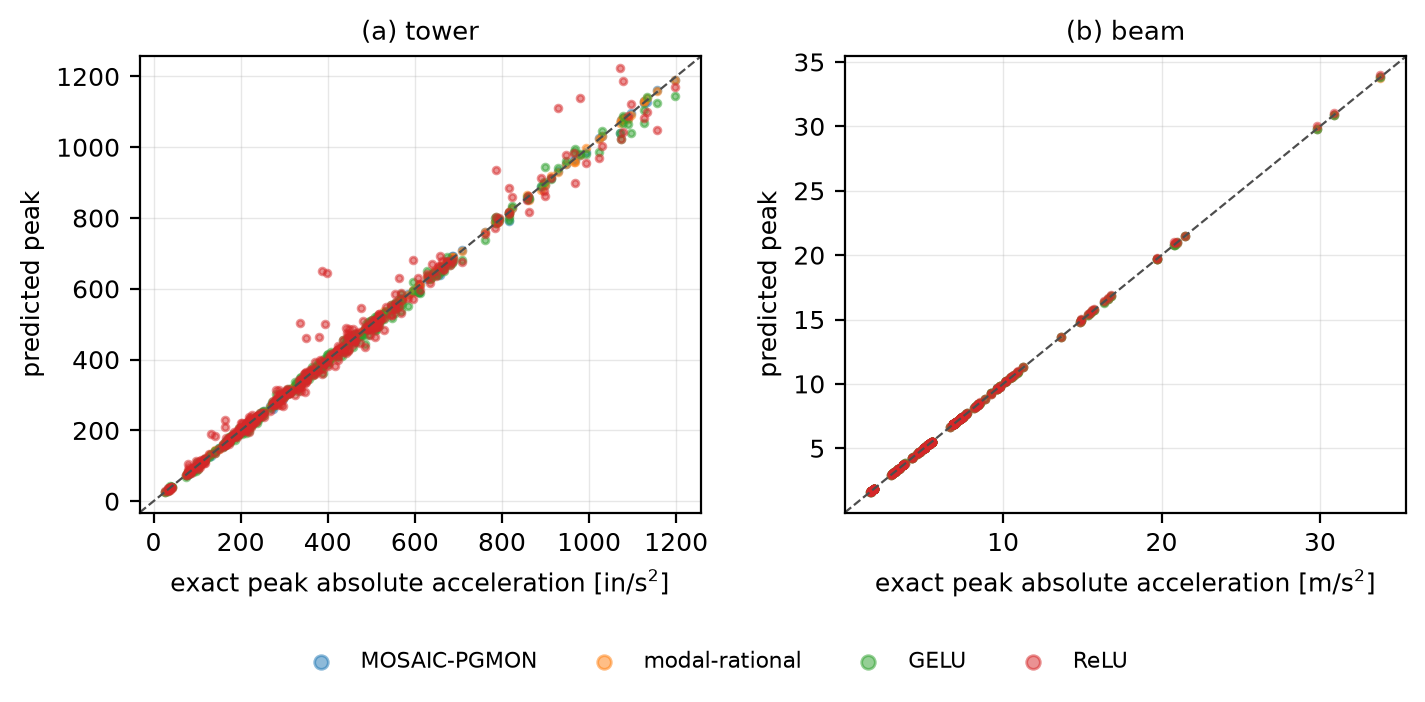}
\caption{Predicted versus reference peak absolute acceleration for the four source configurations: (a) tower and (b) beam. The dashed identity line represents exact peak agreement. Original points and units are preserved, and the legend remains outside the plotting areas. The historical modal-rational legend denotes the Modal Rational row subject to the same implementation association described in Section 3.2.}
\end{figure}

\FloatBarrier
\section{9. Interpreting the tower errors}\label{interpreting-the-tower-errors}

\subsection{9.1. Parameter errors and resonance sensitivity}\label{parameter-errors-and-resonance-sensitivity}

The modal representation makes a useful error analysis possible without modifying a trained network. Let \({\Delta R}_{\text{r}}\) = \({\widehat{R}}_{\text{r}}\) − \(R_{\text{r}}\) and \({\Delta D}_{\text{r}}\) = \({\widehat{D}}_{\text{r}}\) − \(D_{\text{r}}\). Subtracting one predicted modal term from its reference gives the exact identity below. It holds wherever both denominators are nonzero.

\begin{equation}
\frac{{\widehat{\mathbf{R}}}_{r}}{{\widehat{D}}_{r}} - \frac{\mathbf{R}_{r}}{D_{r}} = \frac{\Delta\mathbf{R}_{r}}{{\widehat{D}}_{r}} - \frac{\mathbf{R}_{r}\Delta D_{r}}{D_{r}{\widehat{D}}_{r}}.
\end{equation}

\emph{where} \({\widehat{R}}_{\text{r}}\) is the predicted modal numerator; \(R_{\text{r}}\) is the reference modal numerator; \({\widehat{D}}_{\text{r}}\) is the predicted modal denominator; \(D_{\text{r}}\) is the reference modal denominator; \({\Delta R}_{\text{r}}\) is the numerator difference; \({\Delta D}_{\text{r}}\) is the denominator difference.

\begin{equation}
\parallel \widehat{\mathbf{H}} - \mathbf{H} \parallel_{2} \leq \sum_{r}^{}\left( \frac{\parallel \Delta\mathbf{R}_{r} \parallel_{2}}{{\widehat{d}}_{r}} + \frac{\parallel \mathbf{R}_{r} \parallel_{2}e_{D,r}}{d_{r}{\widehat{d}}_{r}} \right) + \parallel \widehat{\mathbf{T}} - \mathbf{T} \parallel_{2}.
\end{equation}

\emph{where} \(d_{\text{r}}\) is the modulus of D\_r; \({\widehat{d}}_{\text{r}}\) is the modulus of \(\widehat{D}_{r}\); \(e_{\text{D,r}}\) is the modulus of ΔD\_r; \(\widehat{\mathbf{T}}\) is the predicted residual response; \(\mathbf{T}\) is the reference residual response; \(\text{L₂}\) denotes the L₂ norm.

Here \(d_{\text{r}}\) and \({\widehat{d}}_{\text{r}}\) are the moduli of \(D_{\text{r}}\) and \({\widehat{D}}_{\text{r}}\), and \(e_{\text{D,r}}\) is the modulus of \({\Delta D}_{\text{r}}\). This bound separates numerator error, denominator error and the residual response. It is reference-assisted: it uses information about the true modal quantities and is not an online certificate for an unseen design. It also explains why a modest parameter error may be important when a denominator is small. Modal-surrogate and rational-approximation studies provide the relevant representation context {[}3, 10{]}.

\begin{equation}
e_{D,r} \leq \left( 2\omega_{r} + e_{\omega,r} \right)e_{\omega,r} + 2\text{abs}(\nu)\left( {\widehat{\zeta}}_{r}e_{\omega,r} + \omega_{r}e_{\zeta,r} \right).
\end{equation}

\emph{where} \(e_{\text{ω,r}}\) is the absolute natural-frequency error; \(e_{\text{ζ,r}}\) is the absolute damping-ratio error; \(\nu\) is the query angular frequency; \(\text{abs}\) denotes absolute value.

The quantities \(e_{\text{ω,r}}\) and \(e_{\text{ζ,r}}\) denote the absolute errors in natural frequency and damping ratio, respectively; abs(\(\nu\)) is the absolute query frequency. For an isolated lightly damped mode, holding its numerator and damping fixed and evaluating near resonance, the first-order relative sensitivity scales approximately as (1/\(\zeta_{\text{r}}\))\textbar{}\(e_{\text{ω,r}}\)\textbar/\(\omega_{\text{r}}\). At 5\% damping, a 0.1\% natural-frequency perturbation can therefore produce about a 2\% local complex-FRF perturbation. This is an analytical illustration, not a measured diagnosis of the tower network.

The distinction is important for the archived experiment: if MR uses exact supplied frequencies and prescribed damping, its denominator-error term is zero for those anchors. Its remaining error cannot be blamed on learning those same frequencies. Numerator learning, the correction branch, the reduced representation, and the evaluation weighting remain relevant. In a design-only implementation, anchor prediction introduces an additional denominator contribution. PGMON's learned shifts can also modify that contribution even when its initial anchor is exact.

Frequency ordering can create another difficulty. Two modes may exchange order as a design changes, making an indexed numerator target switch abruptly while the summed physical response remains continuous. The beam's directional alignment reduces one such ambiguity, but the tower's frequency-ordered labels do not automatically receive the same protection. This is a plausible mechanism, supported by the mathematics of modal correspondence, not a measured share of the tower error {[}14, 42{]}.

The available comparisons do not isolate one dominant source of tower error. Parameter substitution or exact-modal controls on matched cases would be needed to quantify the mechanisms above. Neither input dimensionality nor retained modal order alone establishes that attribution.

\subsection{9.2. Response weighting, omitted modes and difficult cases}\label{response-weighting-omitted-modes-and-difficult-cases}

A frequency-domain norm and a response-history norm are not interchangeable. With \(\mathbf{E}\) = \(H_{\text{o}}\) − \(H_{\text{o}}\), Parseval's identity gives the following relation for a zero-state response over the full reconstructed window, with the same real-valued filter and Fourier convention:

\begin{equation}
\parallel \mathbf{e}_{k} \parallel_{L_{2}}^{2} = \frac{1}{2\pi}\int_{- \infty}^{\infty}\nu^{2k}w(\nu)^{2} \parallel \mathbf{E}(\nu)\mathbf{A}(\nu) \parallel_{2}^{2}\, d\nu,\quad\quad k = 0,1,2.
\end{equation}

\emph{where} \(\mathbf{e}_{\text{k}}\) is the response error of derivative order k; \(\mathbf{E}\) is the complex FRF error \(\widehat{H}-H\); \(\mathbf{A}\) is the input spectrum; \(w\) is the common real-valued filter; \(k\) equals 0, 1 and 2 for displacement, velocity and relative acceleration.

The cases \(k\) = 0, 1 and 2 correspond to displacement, velocity and relative acceleration. Squared acceleration-error energy contains a fourth-power frequency factor. This does not prove that high frequencies dominate every record: the input spectrum, residues and reference response also matter. It does show why an improvement in an unweighted FRF objective need not improve the same response metric. Multiple input directions must be combined coherently before taking the norm; cross terms cannot generally be discarded.

\begin{equation}
{\widehat{\mathbf{a}}}_{\text{abs}} - \mathbf{a}_{\text{abs}} = {\widehat{\mathbf{a}}}_{\text{rel}} - \mathbf{a}_{\text{rel}}.
\end{equation}

\emph{where} \({\widehat{a}}^{\text{abs}}\) is the predicted absolute acceleration; \(a^{\text{abs}}\) is the reference absolute acceleration; \({\widehat{a}}^{\text{rel}}\) is the predicted relative acceleration; \(a^{\text{rel}}\) is the reference relative acceleration.

Equation (28) assumes the same direct input term in both predictions. Relative percentages can still differ because their reference norms differ. A near-zero response or strong cancellation can produce a large percentage from a small absolute discrepancy. The appropriate report retains both the percentage and the physical amplitude of that case, rather than automatically excluding it.

Modal truncation and learned discrepancy also interact. Let \(y_{\text{N}}\) be the exact response of the retained modal representation, processed identically to the reference \(y_{\text{ref}}\) and the prediction \(\widehat{y}\). Define \(\mathbf{e}_{\text{L}}\) = \(\widehat{y}\) − \(y_{\text{N}}\) and \(\mathbf{e}_{\text{T}}\) = \(y_{\text{N}}\) − \(y_{\text{ref}}\). Then

\begin{equation}
\parallel \widehat{\mathbf{y}} - \mathbf{y}_{\text{ref}} \parallel_{2}^{2} = \parallel \mathbf{e}_{L} \parallel_{2}^{2} + \parallel \mathbf{e}_{T} \parallel_{2}^{2} + 2\text{Re}\langle\mathbf{e}_{L},\mathbf{e}_{T}\rangle.
\end{equation}

\emph{where} \(y_{\text{N}}\) is the exact response of the retained modal representation; \(y_{\text{ref}}\) is the declared reference response; \(\widehat{y}\) is the learned prediction; \(\mathbf{e}_{\text{L}}\) is the learned-model discrepancy; \(\mathbf{e}_{\text{T}}\) is the truncation discrepancy; \(\text{Re}\) denotes the real part; \(\text{⟨}\text{·,·}\text{⟩}\) denotes the complex inner product.

The interaction term may be negative. A learned correction can partially compensate for omitted dynamics, so exact truncation error is not a universal lower bound for every corrected network. Conversely, more retained modes do not guarantee a lower learned error if the additional quantities are harder to estimate. This identity remains valid after a common linear crop; the simple full-window frequency integral in Eq. (27) must not be applied unchanged to a cropped relative metric.

\Needspace{7\baselineskip}
\begin{table}[H]
\tablecaption{What the present evidence supports about the tower.}

\centering
\begin{tabular}{@{}
  >{\raggedright\arraybackslash}p{(\columnwidth - 2\tabcolsep) * \real{0.4762}}
  >{\raggedright\arraybackslash}p{(\columnwidth - 2\tabcolsep) * \real{0.5238}}@{}}
\toprule\noalign{}
\begin{minipage}[b]{\linewidth}\raggedright
\textbf{Supported observation or identity}
\end{minipage} & \begin{minipage}[b]{\linewidth}\raggedright
\textbf{Limit on the interpretation}
\end{minipage} \\
\midrule\noalign{}
Tower response errors exceed the beam's for the reported modal configurations. & The two structures differ in variables, training and evaluation; no single cause is isolated. \\
Derivative response weights spectral discrepancy differently. & This does not identify the error-dominant bands without spectra and case-level data. \\
Static correction restores a zero-frequency (DC) contribution. & It does not restore omitted in-band resonances. \\
Modal and discrepancy errors may cancel. & Replacing one predicted quantity by an exact one is not an additive causal decomposition. \\
\bottomrule\noalign{}
\end{tabular}
\end{table}

\subsection{9.3. Damping sensitivity and scope of physical claims}\label{damping-sensitivity-and-scope-of-physical-claims}

The source reports a paired tower comparison in which the earlier Rayleigh prescription gives approximately 0.57\% mean displacement error and 0.68\% mean absolute-acceleration error for the best configuration, compared with 0.73\% and 1.27\% under constant 5\% damping. These values show that the damping assumption can materially change the measured surrogate error. They do not show that one damping law is physically correct for every structure or that the network itself improved.

The Rayleigh ratios are 5.00\% near 0.8 Hz, 10.02\% at 3 Hz and 22.11\% at 7 Hz. Larger damping suppresses and broadens individual resonant contributions. The response of a multi-mode system can additionally involve cancellation, so the effect on a global relative error must be measured rather than inferred solely from one modal ratio. The modeling cautions in the damping literature support explicit reporting, not the claim that constant damping is the only practical choice {[}28, 29, 30{]}.

For the learned modal core, positive frequencies and damping place the second-order poles in the left half-plane. With real, frequency-independent numerator matrices, this gives a stable real rational realization. A real static remainder is a feedthrough term. Complex constants, unrestricted polynomial tails and neural spectral corrections require separate analysis; positive frequency and damping parameters do not prove passivity, reciprocity or causality of the complete corrected and filtered prediction.

The same distinction applies to ``physics-exact.'' Analytical evaluation of the retained modal formula is exact for its supplied parameters. Predicted parameters can be inaccurate, a clipped or regularized denominator can differ from that formula, and a truncated model can differ from the complete FE dynamics. The present paper uses ``physics-embedded'' to describe the representation, not to erase those approximation levels.

\FloatBarrier
\section{10. Computational cost and reproducibility}\label{computational-cost-and-reproducibility}

A custom-made FEA solver (PRIME-CAE) was developed and verified for this research to automatically produce data, create a database, and run the described models. The code can be provided to the interested reader upon request. Table 9 shows the computational times of the database generation.

\Needspace{7\baselineskip}
\begin{table}[H]
\tablecaption{Source-reported component timings on an Apple-silicon laptop, four central processing unit (CPU) threads. These are not established end-to-end speed-ups for a design-only deployment.}

\centering
\begin{tabular}{@{}
  >{\raggedright\arraybackslash}p{(\columnwidth - 4\tabcolsep) * \real{0.4762}}
  >{\raggedright\arraybackslash}p{(\columnwidth - 4\tabcolsep) * \real{0.2619}}
  >{\raggedright\arraybackslash}p{(\columnwidth - 4\tabcolsep) * \real{0.2619}}@{}}
\toprule\noalign{}
\begin{minipage}[b]{\linewidth}\raggedright
\textbf{Operation}
\end{minipage} & \begin{minipage}[b]{\linewidth}\raggedright
\textbf{Tower}
\end{minipage} & \begin{minipage}[b]{\linewidth}\raggedright
\textbf{Beam}
\end{minipage} \\
\midrule\noalign{}
Assembly and eigenanalysis per design & 56 ms & 1.36 s \\
Modal FRF from stored modal data & 1.2 ms & Not separately reported \\
Network FRF evaluation, all recorded outputs & 28 ms / 1,401 frequencies & 93 ms / 2,001 frequencies \\
Database generation, 20,000 designs & 18.7 min & 7.5 h \\
Training: sum of archived family compute times & 20.7 h & 12.3 h \\
\bottomrule\noalign{}
\end{tabular}
\end{table}

Computational cost depends on the information available at inference. Conditional models that require exact modal frequencies must include the cost of obtaining these quantities unless they are assumed to be precomputed or supplied externally. In the archived tower measurements, evaluation of the analytical modal sum from stored modal data was faster than neural-network evaluation. The previously reported approximate 2X and 15X ratios are therefore not interpreted as validated end-to-end speed-ups because the compared workflows did not use a common inference boundary or account consistently for modal-anchor generation and preprocessing. Accordingly, only component-level timings are reported here. Establishing an end-to-end computational advantage requires a matched timing study that includes modal-information generation, preprocessing, surrogate inference, response reconstruction, and the corresponding reference-analysis costs under the same hardware, batching, and accuracy conditions.

A design-only implementation may have a different cost profile, but its frequency predictor, preprocessing and response reconstruction must be counted. Likewise, differentiating a neural output while holding externally supplied modal frequencies fixed does not give the complete design derivative when those frequencies depend on the design. Accurate gradients and an optimization speed advantage remain claims to validate, not consequences of using automatic differentiation.

The archive identifies fixed population splits, training seeds 20260713 for the tower and 20260918 for the beam, checkpoint configurations, per-epoch histories and response arrays. The numerical tables and restored curves refer to those runs. A filename alone does not establish equivalence between historical implementations.

A reproducible release should associate each result with its model definition, inference inputs, preprocessing statistics, modal set, damping, output ordering, objective, selected epoch and source version. The signal manifest should retain component identifiers, units, sample intervals, scaling and any alignment or filtering assumptions. Raw response arrays and the aggregation script are preferable to summary numbers alone, particularly when near-zero channels or observed maxima matter.

Model-selection decisions must remain distinct from the final test evaluation {[}37{]}. The present ten-design histories do not demonstrate a population-wide failure probability, and one seed does not quantify training variability. These restrictions do not erase the reported accuracy; they specify the extent of the evidence. Further training or broader tests should be reported as additional experiments rather than silently replacing the archived protocol.

\FloatBarrier
\section{11. Conclusions and next steps}\label{conclusions-and-next-steps}

Modal Rational provides the proposed foundation: a neural map estimates structural modal quantities, and an analytical decoder constructs the dynamic response. PGMON extends that formulation with bounded changes to modal parameters and residual flexibility. MOSAIC-MR and MOSAIC-PGMON add bounded spectral discrepancies. The distinction between these methods lies in their forward computations, not in the activation used inside an encoder or the name of a saved software module.

The archived two-structure evaluation supports accurate finite-band response reconstruction under its declared information budget. The recorded modal configurations give mean displacement errors of 0.73--0.76\% for the tower and 0.029--0.116\% for the beam. Acceleration, peak and maximum-error results are reported separately because they do not follow automatically from displacement accuracy. The corrections have a clearer numerical benefit in some beam configurations than in the narrow-band tower comparison; no uniform ranking is claimed.

The benchmark histories are prescribed acceleration inputs, applied through their equivalent inertial forces. Their origin in ground-motion data does not make either example an earthquake-resistant design. No code compliance, seismic safety or site-specific hazard claim is made. The beam is a structural-dynamics benchmark, and its low history errors must not be interpreted as uniform accuracy in weakly excited parts of the FRF band.

The tower--beam difference cannot be reduced to FE model size. The tower varies stiffness and mass over a larger input space, whereas the beam varies ten masses and benefits from modal correspondence preprocessing. The error analysis shows how modal-parameter errors, omitted dynamics and excitation-dependent weighting enter the response. It also shows why static consistency, high correlation or a low mean error does not certify every dynamic channel. The relative importance of these mechanisms still requires matched diagnostics rather than an assumed root cause.

The same physical formulations can be used with finite-dimensional vector mappings and incorporated into neural operators. Future work will develop operator architectures around the proposed MR, PGMON and MOSAIC components, with explicit branch/input representations and output-query coordinates. DeepONet, Fourier and Laplace operator formulations provide relevant foundations {[}7, 8, 23{]}. Their integration will be tested against the standalone methods under matched information, data and computational budgets, rather than assumed to improve accuracy by construction. Operator error analyses and existing reduced-order-informed approaches will guide that evaluation {[}22, 25{]}.

Future comparisons will retain the same connection between formulation, trained computation and evaluated response. This is necessary to distinguish an architectural benefit from a change in input information, damping assumption, or benchmark.

In conclusion, the principal contribution of this study to the structural engineering literature and professional practice is demonstrating that neural surrogate models can achieve significant computational acceleration without abandoning fundamental structural mechanics. By rigorously enforcing analytical frequency-response identities instead of heuristic curve fitting, the proposed hierarchy transitions machine learning from abstract data fitting into an interpretable, physically grounded, and engineering-ready tool for high-rate structural dynamic assessment and design exploration.

\manuscriptheading{Acknowledgements}

The author acknowledges Maltepe University for providing computational resources and institutional support to conduct this research.

\manuscriptheading{CRediT authorship contribution statement}

Amir Nasrollahi and Mehmet Sefa Orak: Conceptualization, Methodology, Software, Validation, Formal analysis, Investigation, Data curation, Writing - original draft, Writing - review \& editing, Visualization, Project administration.

\manuscriptheading{Declaration of competing interests}

The author declares that he has no known competing financial interests or personal relationships that could have appeared to influence the work reported in this paper.

\subsection{Data availability}\label{data-availability}

The structural databases, trained models and evaluation scripts are described in the authors' archive. Access conditions and redistribution permissions for the benchmark input signals must follow the source database terms. Requests for the research artifacts should be directed to the corresponding author.

\subsection{Declaration of artificial intelligence (AI) assistance}\label{declaration-of-artificial-intelligence-ai-assistance}

ChatGPT (OpenAI) assisted with spell checking, grammatical revision, figure layout, and scripts for plots derived from the supplied numerical tables and input-selection archive. No new model-training results or FE responses were generated by that assistance. The authors are responsible for the methods, numerical evidence, references and final submitted text.

\FloatBarrier
\section{Appendix A. Recorded training settings}\label{appendix-a.-recorded-training-settings}

The following settings summarize the source configurations. They are not a prescription for the ongoing training campaign and do not imply that the two structural tasks used identical optimization. Per-family values within a reported range must be recovered from the corresponding checkpoint configuration.

\Needspace{7\baselineskip}
\begin{table}[H]
\tablecaption{Optimization and evaluation settings retained from the archived study.}

\centering
\begin{tabular}{@{}
  >{\raggedright\arraybackslash}p{(\columnwidth - 4\tabcolsep) * \real{0.2778}}
  >{\raggedright\arraybackslash}p{(\columnwidth - 4\tabcolsep) * \real{0.3413}}
  >{\raggedright\arraybackslash}p{(\columnwidth - 4\tabcolsep) * \real{0.3810}}@{}}
\toprule\noalign{}
\begin{minipage}[b]{\linewidth}\raggedright
\textbf{Setting}
\end{minipage} & \begin{minipage}[b]{\linewidth}\raggedright
\textbf{Tower}
\end{minipage} & \begin{minipage}[b]{\linewidth}\raggedright
\textbf{Beam}
\end{minipage} \\
\midrule\noalign{}
Optimizer & AdamW; β₁ = 0.9, β₂ = 0.95; ε = 10⁻⁷ & Same optimizer constants \\
Epochs / updates per epoch & 300 / 300 & 40--100 / 300 \\
Batch & 512 design--frequency pairs & 64 designs × 8 frequencies \\
Learning rate & Peak 5 × 10⁻⁵; minimum 10⁻⁷ & Peaks 10⁻³ / 10⁻⁴ / 3 × 10⁻⁴; minimum 10⁻⁶ \\
Warm-up and decay & 50 warm-up epochs; cosine decay & 2--10 warm-up epochs; cosine decay \\
Weight decay / dropout & 0.01 / 0.03 & 10⁻⁴ with stated exclusions / 0.03 \\
Gradient clipping & Global norm 0.5 & Global norm 2.0 \\
Weight averaging & None & EMA 0.999 from 15\% of training \\
Curriculum & All declared terms from epoch 1 & Stages at 0\%, 8\%, 25\% and 55\% \\
Frequency draws & 60\% resonance; 40\% uniform & 38\% resonance; 12\% antiresonance; 12\% uniform; 8\% log; 5\% edge; 25\% response-energy \\
Main data weights & MSE 1; complex 0.10; magnitude 0.05; phase 0.01 & MSE 1; complex 0.05; magnitude 0.10; phase 0.05; derivative 0.05; peak 0.05; time 0.25 \\
Validation & 80 fixed batches & 80 fixed batches; response check on 24 designs and 2 inputs every 5 epochs \\
Selection & Composite validation; gap-score diagnostic & Validation response; gap-score diagnostic \\
Training seed & 20260713 & 20260918 \\
\bottomrule\noalign{}
\end{tabular}
\end{table}

The beam source reports a training-target standard-deviation floor of 10⁻⁵ and an inactive-output threshold of 3 × 10⁻⁵ times the largest target standard deviation. It reports six masked transfer entries out of nine. The precise use of that mask in the objective, audit and reconstructed histories must be documented with the results, rather than inferred from the output count.

The archived modal corrections use a 3\% frequency bound and damping in {[}0.005, 0.30{]}. Auxiliary tower weights are 1.0 for residue supervision, 0.25 for the frequency prior, 0.02 for damping and 0.02 for flexibility shrinkage. The beam records additional correction, residual-pole, gate and router penalties. The publication should retain their checkpoint-specific definitions rather than claim that all eight labels used one identical objective.

Figures 2--7 and 10--12 retain the supplied structural and numerical visual evidence. Figure 9 is redrawn from the unchanged response-error tables. Figure 8 is reconstructed from the uploaded spectrum catalog and the verified selection factors; it is a benchmark-selection plot, not an acceleration-time-history reconstruction. Enlarging an embedded source plot does not create new numerical resolution. No training trajectory, response trace or scatter point was synthesized to fill a missing result.

\FloatBarrier
\FloatBarrier
\section{References}\label{references}

\href{https://www.wiley-vch.de/de/fachgebiete/ingenieurwesen/modal-testing-978-0-86380-218-8}{{[}1{]} D.J. Ewins, Modal Testing: Theory, Practice and Application, second ed., Research Studies Press, Baldock, 2000. ISBN 9780863802188.}

\href{https://www.sciencedirect.com/science/article/pii/S0045794909000066}{{[}2{]} L. Pichler, H.J. Pradlwarter, G.I. Schuëller, A mode-based meta-model for the frequency response functions of uncertain structural systems, Computers \& Structures 87 (2009) 332--341. doi: 10.1016/j.compstruc.2008.12.013.}

\href{https://research.chalmers.se/en/publication/524891}{{[}3{]} M. Gibanica, T.J.S. Abrahamsson, Data-driven modal surrogate model for frequency response uncertainty propagation, Probabilistic Engineering Mechanics 66 (2021) 103142. doi: 10.1016/j.probengmech.2021.103142.}

\href{https://epubs.siam.org/doi/10.1137/130932715}{{[}4{]} P. Benner, S. Gugercin, K. Willcox, A survey of projection-based model reduction methods for parametric dynamical systems, SIAM Review 57 (2015) 483--531. doi: 10.1137/130932715.}

\href{https://doi.org/10.1016/j.jcp.2018.10.045}{{[}5{]} M. Raissi, P. Perdikaris, G.E. Karniadakis, Physics-informed neural networks: A deep learning framework for solving forward and inverse problems involving nonlinear partial differential equations, Journal of Computational Physics 378 (2019) 686--707. doi: 10.1016/j.jcp.2018.10.045.}

\href{https://www.nature.com/articles/s42254-021-00314-5}{{[}6{]} G.E. Karniadakis, I.G. Kevrekidis, L. Lu, P. Perdikaris, S. Wang, L. Yang, Physics-informed machine learning, Nature Reviews Physics 3 (2021) 422--440. doi: 10.1038/s42254-021-00314-5.}

\href{https://www.nature.com/articles/s42256-021-00302-5}{{[}7{]} L. Lu, P. Jin, G. Pang, Z. Zhang, G.E. Karniadakis, Learning nonlinear operators via DeepONet based on the universal approximation theorem of operators, Nature Machine Intelligence 3 (2021) 218--229. doi: 10.1038/s42256-021-00302-5.}

\href{https://arxiv.org/abs/2010.08895}{{[}8{]} Z. Li, N. Kovachki, K. Azizzadenesheli, B. Liu, K. Bhattacharya, A. Stuart, A. Anandkumar, Fourier neural operator for parametric partial differential equations, International Conference on Learning Representations, 2021. arXiv:2010.08895.}

\href{https://www.jmlr.org/papers/v24/21-1524.html}{{[}9{]} N. Kovachki, Z. Li, B. Liu, K. Azizzadenesheli, K. Bhattacharya, A. Stuart, A. Anandkumar, Neural operator: Learning maps between function spaces with applications to PDEs, Journal of Machine Learning Research 24 (89) (2023) 1--97.}

\href{https://www.sintef.no/en/software/vector-fitting/}{{[}10{]} B. Gustavsen, A. Semlyen, Rational approximation of frequency domain responses by vector fitting, IEEE Transactions on Power Delivery 14 (1999) 1052--1061. doi: 10.1109/61.772353.}

\href{https://onlinelibrary.wiley.com/doi/10.1002/pamm.202300091}{{[}11{]} J. Schultz, J. van Delden, C. Blech, S.C. Langer, T. Lüddecke, Deep learning for frequency response prediction of a multimass oscillator, Proceedings in Applied Mathematics and Mechanics 23 (2023) e202300091. doi: 10.1002/pamm.202300091.}

\href{https://proceedings.neurips.cc/paper_files/paper/2024/hash/bbd7d8bd780fcf7143add2317ba04638-Abstract-Conference.html}{{[}12{]} J. van Delden, J. Schultz, C. Blech, S.C. Langer, T. Lüddecke, Learning to predict structural vibrations, Advances in Neural Information Processing Systems 37 (2024).}

\href{https://doi.org/10.1016/j.engstruct.2020.110704}{{[}13{]} R. Zhang, Y. Liu, H. Sun, Physics-guided convolutional neural network (PhyCNN) for data-driven seismic response modeling, Engineering Structures 215 (2020) 110704. doi: 10.1016/j.engstruct.2020.110704.}

\href{https://link.springer.com/article/10.1007/s00366-026-02310-8}{{[}14{]} C. Lin, J. Ding, J. Chen, Y. Wang, J. Jiao, P.-I. Mak, N. Wang, Graph-based surrogate modeling for structural eigenvalue problems with modal matching training, Engineering with Computers 42 (2026) 72. doi: 10.1007/s00366-026-02310-8.}

\href{https://epubs.siam.org/doi/10.1137/19M1274067}{{[}15{]} L. Lu, X. Meng, Z. Mao, G.E. Karniadakis, DeepXDE: A deep learning library for solving differential equations, SIAM Review 63 (2021) 208--228. doi: 10.1137/19M1274067.}

\href{https://doi.org/10.1016/j.cma.2021.113741}{{[}16{]} E. Haghighat, M. Raissi, A. Moure, H. Gomez, R. Juanes, A physics-informed deep learning framework for inversion and surrogate modeling in solid mechanics, Computer Methods in Applied Mechanics and Engineering 379 (2021) 113741. doi: 10.1016/j.cma.2021.113741.}

\href{https://epubs.siam.org/doi/10.1137/20M1318043}{{[}17{]} S. Wang, Y. Teng, P. Perdikaris, Understanding and mitigating gradient flow pathologies in physics-informed neural networks, SIAM Journal on Scientific Computing 43 (2021) A3055--A3081. doi: 10.1137/20M1318043.}

\href{https://arxiv.org/abs/2006.09661}{{[}18{]} V. Sitzmann, J.N.P. Martel, A.W. Bergman, D.B. Lindell, G. Wetzstein, Implicit neural representations with periodic activation functions, Advances in Neural Information Processing Systems 33 (2020) 7462--7473.}

\href{https://arxiv.org/abs/2006.10739}{{[}19{]} M. Tancik, P.P. Srinivasan, B. Mildenhall, S. Fridovich-Keil, N. Raghavan, U. Singhal, R. Ramamoorthi, J.T. Barron, R. Ng, Fourier features let networks learn high frequency functions in low dimensional domains, Advances in Neural Information Processing Systems 33 (2020) 7537--7547.}

\href{https://proceedings.mlr.press/v97/rahaman19a.html}{{[}20{]} N. Rahaman, A. Baratin, D. Arpit, F. Draxler, M. Lin, F. Hamprecht, Y. Bengio, A. Courville, On the spectral bias of neural networks, Proceedings of Machine Learning Research 97 (2019) 5301--5310.}

\href{https://doi.org/10.1016/j.cma.2022.114778}{{[}21{]} L. Lu, X. Meng, S. Cai, Z. Mao, S. Goswami, Z. Zhang, G.E. Karniadakis, A comprehensive and fair comparison of two neural operators (with practical extensions) based on FAIR data, Computer Methods in Applied Mechanics and Engineering 393 (2022) 114778. doi: 10.1016/j.cma.2022.114778.}

\href{https://doi.org/10.1093/imatrm/tnac001}{{[}22{]} S. Lanthaler, S. Mishra, G.E. Karniadakis, Error estimates for DeepONets: A deep learning framework in infinite dimensions, Transactions of Mathematics and Its Applications 6 (2022) tnac001. doi: 10.1093/imatrm/tnac001.}

\href{https://www.nature.com/articles/s42256-024-00844-4}{{[}23{]} Q. Cao, S. Goswami, G.E. Karniadakis, Laplace neural operator for solving differential equations, Nature Machine Intelligence 6 (2024) 631--640. doi: 10.1038/s42256-024-00844-4.}

\href{https://link.springer.com/article/10.1007/s10915-021-01462-7}{{[}24{]} S. Fresca, L. Dedè, A. Manzoni, A comprehensive deep learning-based approach to reduced order modeling of nonlinear time-dependent parametrized PDEs, Journal of Scientific Computing 87 (2021) 61. doi: 10.1007/s10915-021-01462-7.}

\href{https://pubmed.ncbi.nlm.nih.gov/40174356/}{{[}25{]} L.-H. Yang, X.-L. Luo, Z.-B. Yang, C.-F. Nan, X.-F. Chen, Y. Sun, FE reduced-order model-informed neural operator for structural dynamic response prediction, Neural Networks 188 (2025) 107437. doi: 10.1016/j.neunet.2025.107437.}

\href{https://proceedings.mlr.press/v15/glorot11a.html}{{[}26{]} X. Glorot, A. Bordes, Y. Bengio, Deep sparse rectifier neural networks, Proceedings of Machine Learning Research 15 (2011) 315--323.}

\href{https://arxiv.org/abs/1606.08415}{{[}27{]} D. Hendrycks, K. Gimpel, Gaussian error linear units (GELUs), arXiv preprint arXiv:1606.08415, 2016.}

\href{https://onlinelibrary.wiley.com/doi/10.1002/eqe.541}{{[}28{]} J.F. Hall, Problems encountered from the use (or misuse) of Rayleigh damping, Earthquake Engineering \& Structural Dynamics 35 (2006) 525--545. doi: 10.1002/eqe.541.}

\href{https://doi.org/10.1061/(ASCE)0733-9445(2008)134:4(581)}{{[}29{]} F.A. Charney, Unintended consequences of modeling damping in structures, Journal of Structural Engineering 134 (2008) 581--592. doi: 10.1061/(ASCE)0733-9445(2008)134:4(581).}

\href{https://onlinelibrary.wiley.com/doi/10.1002/eqe.2357}{{[}30{]} P. Jehel, P. Léger, A. Ibrahimbegovic, Initial versus tangent stiffness-based Rayleigh damping in inelastic time history seismic analyses, Earthquake Engineering \& Structural Dynamics 43 (2014) 467--484. doi: 10.1002/eqe.2357.}

\href{https://openaccess.thecvf.com/content_cvpr_2016/html/He_Deep_Residual_Learning_CVPR_2016_paper.html}{{[}31{]} K. He, X. Zhang, S. Ren, J. Sun, Deep residual learning for image recognition, Proceedings of the IEEE Conference on Computer Vision and Pattern Recognition, 2016, pp. 770--778. doi: 10.1109/CVPR.2016.90.}

\href{https://ojs.aaai.org/index.php/AAAI/article/view/11671}{{[}32{]} E. Perez, F. Strub, H. de Vries, V. Dumoulin, A. Courville, FiLM: Visual reasoning with a general conditioning layer, Proceedings of the AAAI Conference on Artificial Intelligence 32 (2018) 3942--3951. doi: 10.1609/aaai.v32i1.11671.}

\href{https://arxiv.org/abs/1706.03762}{{[}33{]} A. Vaswani, N. Shazeer, N. Parmar, J. Uszkoreit, L. Jones, A.N. Gomez, Ł. Kaiser, I. Polosukhin, Attention is all you need, Advances in Neural Information Processing Systems 30 (2017) 5998--6008.}

\href{https://arxiv.org/abs/1412.6980}{{[}34{]} D.P. Kingma, J. Ba, Adam: A method for stochastic optimization, International Conference on Learning Representations, 2015. arXiv:1412.6980.}

\href{https://arxiv.org/abs/1711.05101}{{[}35{]} I. Loshchilov, F. Hutter, Decoupled weight decay regularization, International Conference on Learning Representations, 2019. arXiv:1711.05101.}

\href{https://www.jmlr.org/papers/v15/srivastava14a.html}{{[}36{]} N. Srivastava, G. Hinton, A. Krizhevsky, I. Sutskever, R. Salakhutdinov, Dropout: A simple way to prevent neural networks from overfitting, Journal of Machine Learning Research 15 (2014) 1929--1958.}

\href{https://www.jmlr.org/papers/v11/cawley10a.html}{{[}37{]} G.C. Cawley, N.L.C. Talbot, On over-fitting in model selection and subsequent selection bias in performance evaluation, Journal of Machine Learning Research 11 (2010) 2079--2107.}

\href{https://doi.org/10.1108/eb023562}{{[}38{]} E.N. Dvorkin, K.-J. Bathe, A continuum mechanics based four-node shell element for general non-linear analysis, Engineering Computations 1 (1984) 77--88. doi: 10.1108/eb023562.}

\href{https://onlinelibrary.wiley.com/doi/10.1002/nme.1620220312}{{[}39{]} K.-J. Bathe, E.N. Dvorkin, A formulation of general shell elements---the use of mixed interpolation of tensorial components, International Journal for Numerical Methods in Engineering 22 (1986) 697--722. doi: 10.1002/nme.1620220312.}

\href{https://doi.org/10.1090/S0025-5718-1980-0583502-2}{{[}40{]} T. Ericsson, A. Ruhe, The spectral transformation Lanczos method for the numerical solution of large sparse generalized symmetric eigenvalue problems, Mathematics of Computation 35 (1980) 1251--1268. doi: 10.1090/S0025-5718-1980-0583502-2.}

\href{https://epubs.siam.org/doi/book/10.1137/1.9780898719628}{{[}41{]} R.B. Lehoucq, D.C. Sorensen, C. Yang, ARPACK Users' Guide: Solution of Large-Scale Eigenvalue Problems with Implicitly Restarted Arnoldi Methods, SIAM, Philadelphia, 1998. doi: 10.1137/1.9780898719628.}

\href{https://www.sandv.com/downloads/0308alle.pdf}{{[}42{]} R.J. Allemang, The modal assurance criterion---twenty years of use and abuse, Sound and Vibration 37 (8) (2003) 14--23.}

\href{https://journals.sagepub.com/doi/10.1193/070913EQS197M}{{[}43{]} T.D. Ancheta, R.B. Darragh, J.P. Stewart, E. Seyhan, W.J. Silva, B.S.-J. Chiou, K.E. Wooddell, R.W. Graves, A.R. Kottke, D.M. Boore, T. Kishida, J.L. Donahue, NGA-West2 database, Earthquake Spectra 30 (2014) 989--1005. doi: 10.1193/070913EQS197M.}

\href{https://www.inderscienceonline.com/doi/abs/10.1504/IJMISSP.2018.092929}{{[}44{]} A. Behnam, A. Nasrollahi, Optimal selecting and scaling the earthquake accelerograms according to Eurocode 8 using ranked particles optimisation, International Journal of Machine Intelligence and Sensory Signal Processing 2 (2018) 121--139. doi: 10.1504/IJMISSP.2018.092929.}

\href{https://doi.org/10.1061/JMCEA3.0000098}{{[}45{]} N.M. Newmark, A method of computation for structural dynamics, Journal of the Engineering Mechanics Division 85 (1959) 67--94. doi: 10.1061/JMCEA3.0000098.}

\FloatBarrier
\end{document}